\documentclass[preprintnumbers,amsmath,amssymb,floatfix,11pt,prd,onecolumn,superscriptaddress,nofootinbib]{revtex4}
\usepackage{amsmath,amsfonts,dsfont,amsthm,amssymb}
\usepackage{caption}%or \usepackage[format=hang]{caption}
\usepackage{mathtools}
\usepackage{pdflscape}
\usepackage{multirow}
\usepackage{graphicx}
\usepackage{color}
\usepackage{epsfig}
\usepackage{verbatim}
\usepackage{float}
\usepackage{hyperref}%This option gives active links in colored boxes. OR Use option below
\theoremstyle{plain}

\theoremstyle{definition}

\usepackage{subfigure}
\theoremstyle{remark}

\numberwithin{equation}{section}

\begin{document}
\title{\bf Optical Appearance of a Rotating Black Hole in Nonlinear Electrodynamics Surrounded by Thin Accretion Disks}

\altaffiliation{linzch24@cqu.edu.cn}

\author{Abdul Malik Sultan}
\altaffiliation{ams@uo.edu.pk}
\affiliation{Department of Mathematics, University of Okara, Okara-56300 Pakistan}

\author{Manahil Ali}
\altaffiliation{manomanahil600@gmail.com}
\affiliation{Department of Mathematics, University of Okara, Okara-56300 Pakistan}

\author{Muhammad Israr Aslam}
\altaffiliation{mrisraraslam@gmail.com, israr.aslam@umt.edu.pk}
\affiliation{Department of Mathematics, School of Science, University of Management and Technology, Lahore-$54770$, Pakistan.}

\author{Zi-Chao Lin}
\affiliation{Physics Department, Chongqing University, Chongqing 401331, China}

\begin{abstract}
 This work investigates the optical appearance of a rotating black hole (BH) in nonlinear electrodynamics (NED) using two illumination models, namely a celestial sphere and a thin accretion disk. The BH images are constructed using a backward ray-tracing method together with a fisheye camera model. We examine the effects of the electric charge $Q$ and the NED parameter $\beta$ on the event horizon, shadow, photon ring, and optical appearance for both prograde and retrograde accretion flows. The results indicate that an increase in $\beta$ leads to a larger shadow radius with reduced distortion, whereas increasing $Q$ decreases the shadow size and enhances its deformation. These features are further quantified through the shadow radius and distortion parameter. We also analyze the direct and lensed images of the thin accretion disk, together with their corresponding redshift distributions and emission bands. The redshifted emission is found to dominate the observed images, while the blueshifted region is confined to the vicinity of the photon ring. Our results demonstrate that both $Q$ and $\beta$ leave distinct signatures on the optical appearance of rotating NED BHs, providing useful insights for future high-resolution observations. 
\end{abstract}
\date{\today}
\maketitle

\section{Introduction}

General Relativity (GR) is the prevailing theory of gravitation, describing gravity as a manifestation of the curvature of spacetime generated by the distribution of mass and energy \cite{1}. Among its most remarkable predictions are BHs, which are compact astrophysical objects with gravitational fields so intense that no matter or radiation can escape once it crosses the event horizon. The propagation of light in the vicinity of a BH is strongly influenced by the underlying spacetime geometry, causing phenomena such as gravitational lensing, photon capture, and the formation of BH shadows. The first observational confirmation of light deflection by gravity was provided during the 1919 solar eclipse expedition led by Arthur Eddington, offering one of the earliest experimental validations of GR. More recently, advances in observational astronomy have enabled direct investigations of strong-field gravity through BH imaging, providing new opportunities to test gravitational theories beyond the weak-field regime \cite{2}.

Black holes have remained one of the central topics in gravitational physics since the discovery of the first exact solution of Einstein's field equations. The Schwarzschild solution provides the simplest description of a static, spherically symmetric, and electrically neutral BH \cite{3}. This solution was later generalized to include electric charge, leading to the Reissner-Nordström BH \cite{4,5}. The inclusion of angular momentum gave rise to the Kerr solution, which describes an uncharged rotating BH and plays a fundamental role in relativistic astrophysics because most astrophysical BHs are expected to possess nonzero spin \cite{6}. Subsequently, the Kerr-Newman metric was derived as the most general stationary solution of the Einstein-Maxwell equations, characterized by mass, electric charge, and rotation \cite{7}. Rotating BHs are among the most important solutions of Einstein's field equations because astrophysical BHs are generally expected to possess angular momentum acquired during stellar collapse and subsequent accretion processes \cite{8,9}. Unlike static BHs, rotation gives rise to distinctive relativistic effects, including frame dragging, the formation of an ergoregion, and significant modifications to the innermost stable circular orbit (ISCO), all of which strongly influence the motion of particles and photons in the vicinity of the event horizon \cite{10}. Owing to these unique characteristics, rotating BHs provide a realistic framework for investigating a broad range of astrophysical phenomena, including accretion disk dynamics, gravitational lensing, BH shadows, and electromagnetic emission from compact objects \cite{11,12}. Consequently, Kerr and Kerr-like spacetimes have been extensively employed to study the observational signatures of strong gravitational fields and to test the predictions of General Relativity through high-resolution astronomical observations and theoretical analyses \cite{13}.

Despite the remarkable success of GR and Einstein-Maxwell theory in describing the gravitational and electromagnetic properties of BH, classical charged BH solutions still suffer from spacetime singularities and divergent electromagnetic fields near the central region. In particular, the electric field associated with a point charge becomes unbounded at the origin, resulting in an infinite self-energy within Maxwell's electrodynamics. To overcome this limitation, Born and Infeld introduced a nonlinear extension of classical electrodynamics that regularizes the electromagnetic field in the strong-field regime while recovering Maxwell's theory in the weak-field limit \cite{14}. Subsequently, NED has received considerable attention, as many of its models naturally emerge in string-inspired theories and other fundamental frameworks of high-energy physics \cite{15,16}. The coupling of NED with GR has led to a wide class of charged and regular BH solutions, whose geometric, thermodynamical, and optical properties have been extensively explored in recent years \cite{17,18,18a,19,20,21,22,23,24,25,26,27}. These developments demonstrate that NED provides an effective framework for investigating strong gravitational phenomena and offers a promising avenue for constructing physically viable BH spacetimes beyond the standard Einstein-Maxwell description.

In most recent studies, Bokulic et al. \cite{28} investigated BH mechanics within the framework of NED. They generalized the first law of BH thermodynamics, discussed the corresponding Smarr relation, and analyzed the influence of NED fields on the thermodynamic properties of BHs. Zubair et al. \cite{29} explored the optical appearance of a rotating NED BH generated via the modified Newman-Janis algorithm. Their analysis of photon motion, shadow observables, and energy emission revealed that the NED parameter, along with the BH spin and charge, induces noticeable modifications to the spacetime's optical characteristics. In \cite{30}, the authors investigated the optical properties of a rotating regular BH within the framework of NED field. Their analysis included photon motion, BH shadows, and Hawking radiation, together with constraints on the model parameters using the observational data of M$87^\ast$ and SgrA$^\ast$.  Róis et al. \cite{31} explored the BH solutions in $f(R,T)$ gravity coupled to NED and investigated the role of nonlinear electromagnetic effects on the structure of spacetime. 
Liang et al. \cite{32} explored the observational properties of a charged NED BH with logarithmic corrections. By analyzing quasinormal modes and graybody factors, they demonstrated that dynamical and radiative observables provide an effective way to distinguish BH models with degenerate shadow images. 

In many astrophysical systems, BHs are surrounded by accreting matter that gradually settles into a geometrically thin and optically thick disk. The pioneering model proposed by Shakura and Sunyaev \cite{33} provided the first theoretical description of such accretion disks within the Newtonian framework, while Novikov and Thorne \cite{34} later extended this formulation to GR. In the relativistic thin-disk model, the accreting matter follows nearly circular Keplerian orbits and continuously loses angular momentum through viscous processes, leading to the release of thermal radiation from the disk surface. The emitted flux, radiative efficiency, and spectral properties are closely linked to the underlying spacetime geometry, making thin accretion disks a powerful probe for investigating the observational signatures of BHs and testing gravitational theories in the strong-field regime \cite{35}.

The Event Horizon Telescope has provided the first horizon-scale images of the supermassive BHs $M87^\ast$ and SgrA$^\ast$ \cite{27a,27b,27c,27d}. These observations reveal a bright photon ring surrounding a central dark shadow produced by strong gravitational lensing near the event horizon. Such images offer a powerful way to probe the spacetime geometry and constrain the physical properties of BHs. In this regard, the investigation of the optical appearance of BHs has become an important approach to understanding how light behaves in the strong-field regime of gravity. Among the various theoretical techniques, celestial light-source models provide an effective framework for studying the propagation of photons and the formation of observable images around compact objects under uniform background illumination \cite{36,37,38}. Numerous studies have shown that the observed optical signatures are highly sensitive to the properties of the underlying spacetime, with parameters such as BH rotation, observer inclination, and surrounding electromagnetic fields producing noticeable changes in the shadow boundary and image morphology \cite{39,40}. In recent years, considerable attention has also been devoted to the optical characteristics of thin and thick accretion disks, rotating BH shadows, polarized images, and accretion phenomena in both GR and modified theories of gravity \cite{41,42,43,45,47,48,49,49b,49c}. These investigations demonstrate that optical observables provide a powerful means of probing the geometry of BH spacetimes and exploring possible deviations from classical gravitational predictions. Nevertheless, the appearance of rotating BH arising from NED, particularly in the presence of geometrically thin accretion disks, has received comparatively less attention. A detailed investigation of these systems can therefore provide further insight into how NED influences photon trajectories and the observable properties of strong gravitational fields.

Motivated by the theoretical and observational developments discussed above, in this work we investigate the optical appearance of a rotating BH in a NED field. To achieve this, we employ a backward ray-tracing technique to study the propagation of photons in the BH spacetime and construct the corresponding observable images. Two physically relevant illumination scenarios, namely a celestial light source and a thin accretion disk, are considered to examine different aspects of the optical appearance. We further analyze the influence of the BH spin , electric charge, and NED parameter on the innermost stable circular orbit (ISCO), disk emission, and the resulting shadow and lensing features. Through this analysis, we aim to provide a deeper understanding of the observational signatures of rotating NED BHs and to offer a useful framework for future high-resolution astronomical observations.

The remainder of this paper is organized as follows. In Section {\bf 2}, we define the rotating BH in NED field and derive the fundamental equations governing photon and particle motion, including the circular geodesics and the ISCO properties. In Section {\bf 3}, we investigate the optical appearance of the BH under illumination from a celestial light source using a backward ray-tracing technique. In Section {\bf 4}, we describe the thin accretion disk model and analyze the influence of the BH spin $a$, electric charge $Q$, and NED parameter $\beta$ on the ISCO radius, energy flux, emission spectrum, redshift distribution, and the resulting observable images. Finally, Section {\bf 5} contains the summary and concluding remarks.
\section{Rotating Black Hole Solution in Nonlinear Electrodynamics}
The rotating BH solution in NED employed in this work is adopted from \cite{50} by applying the modified Newman-Janis algorithm to the corresponding static NED spacetime \cite{51,52}. Since the complete derivation of the metric and its associated geometric properties has already been presented in detail in Ref. \cite{29}, we do not repeat those calculations here. Instead, we briefly outline the essential features of the spacetime that are required for the subsequent analysis. The rotating NED BH provides a generalized description of a rotating charged compact object, where nonlinear electromagnetic corrections modify the spacetime geometry in comparison with the standard Kerr or Kerr-Newman solutions. The geometry is characterized by four fundamental parameters, such as the BH mass $M$, spin parameter $a$, electric charge $Q$, and the NED parameter $\beta$, which governs the strength of the nonlinear electromagnetic effects. These parameters determine the horizon structure, photon dynamics, and the corresponding optical properties of the spacetime. For completeness, we summarize below the relevant metric functions and geometrical quantities that form the basis of our analysis of BH shadows, lensing effects, and thin accretion disk images. In this perspective, the rotating BH solution with NED field is defined as \cite{29}
\begin{eqnarray}\nonumber
    ds^{2} &=&
-\frac{\Delta-a^{2}\sin^{2}\theta}{\rho^{2}}\,dt^{2}
+\frac{\rho^{2}}{\Delta}\,dr^{2}
+\rho^{2}\,d\theta^{2} 
+\frac{\sin^{2}\theta}{\rho^{2}}
\big[(r^{2}+a^{2})^{2}
-\Delta a^{2}\sin^{2}\theta\big]d\phi^{2} 
\\\label{1}&+&\frac{2a\sin^{2}\theta}{\rho^{2}}
(\Delta-r^{2}-a^{2})dt\,d\phi,
\end{eqnarray}
where
\begin{eqnarray}\label{2}
 \Delta &=& r^2-2 M r+a^2 +Q^2-\frac{C^2 \kappa^2}{2}+\frac{16 C^{3/2}\kappa^2}{16 \beta^{1/4}}r, \\ \label{3}
  \rho^2&=&r^2 +a^2 \cos^2 \theta,
\end{eqnarray}
in which $C$ denotes a dimensionless integration constant, while $\kappa$ represents the gravitational coupling constant, and we fixed $C=\kappa=1$ throughout the subsequent analysis. The parameter $C$ has no direct physical interpretation.  However, it influences spacetime geometry, and when $C=0$, the solution reduces to the Reissner-Nordström BH \cite{50}. Moreover, in the limit $\beta \rightarrow{\infty}$, the NED corrections formally disappear. However, this limit is not physically meaningful because the corresponding NED Lagrangian vanishes \cite{53}. Therefore, in this work, only finite positive values of $\beta$ are considered. In particular, we restrict our analysis to the ranges $0<\beta \le 2$ and $0<Q<1$, which enable us to systematically examine the influence of NED and electric charge on the properties of the rotating BH. Furthermore, in the absence of electric charge $Q$, the rotating NED solution reduces to the Kerr geometry, while for both $Q=0$ and $a=0$, it further simplifies to the Schwarzschild spacetime.
\begin{figure}[H]
\centering
\subfigure[\tiny][~$Q=0.1$]{\label{a}\includegraphics[width=5.0cm,height=5cm]{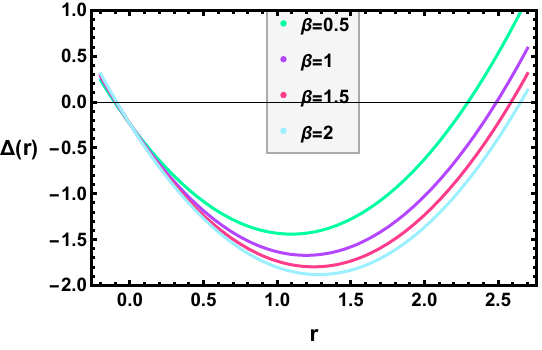}}
\subfigure[\tiny][~$Q=0.5$]{\label{b}\includegraphics[width=5.0cm,height=5cm]{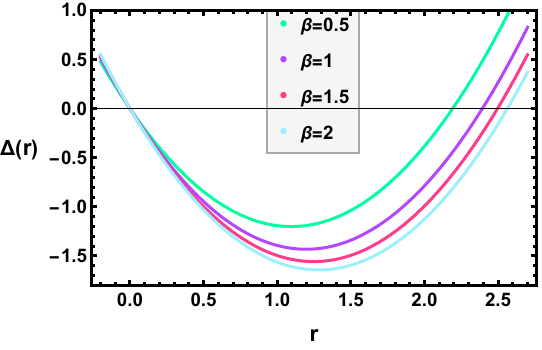}}
\subfigure[\tiny][~$Q=0.9$]{\label{c}\includegraphics[width=5.0cm,height=5cm]{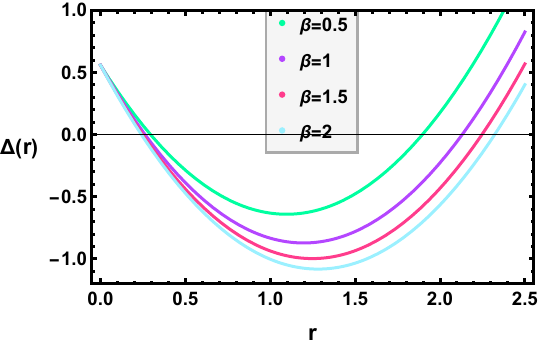}}
\caption{The behavior of the event horizons of the rotating NED BH for different values of $\beta$ and $Q$ with fixed $a=0.5$.}\label{prd1}
\end{figure}
Figure \textbf{\ref{prd1}} shows the radial profile of the metric function $\Delta(r)$ for several values of $\beta$ with fixed spin parameter $a=0.5$. The three panels correspond to fixed electric charge values, namely $Q=0.1,~0.5$ and $0.9$. The intersections of $\Delta(r)$ with the horizontal axis determine the horizon radii. In all cases, two horizons are observed: the inner horizon $r_{in}$ and the outer horizon $r_{out}$, satisfying $r_{in}< r_{out}$. For a fixed value of $Q$, increasing $\beta$ modifies the profile of the metric function. As a result, the points where $\Delta(r)=0$ shift, leading to corresponding changes in the locations of both the inner and outer horizons. It is also observed that the separation between the curves corresponding to different values of $\beta$ gradually decreases as the electric charge increases from $Q=0.1$ to $Q=0.9$. This shows that the NED parameter $\beta$ has a significant influence on the horizon structure of the rotating NED BH.

The optical appearance of a BH is determined by the trajectories of photons propagating through the surrounding spacetime. In the presence of a thin accretion disk, photons emitted by the disk experience a strong gravitational deflection before reaching a distant observer. Those passing close to the photon sphere follow unstable null geodesics, where some are captured by the BH, while others escape to infinity, giving rise to the characteristic dark shadow enclosed by a bright boundary. Therefore, the shadow structure and the emission from the accretion disk provide valuable information about the underlying spacetime geometry and the influence of NED. The motion of photons is governed by the Hamilton-Jacobi equation, which can be written as \cite{54}
\begin{eqnarray} \label{4}
   \frac{\partial I}{\partial \sigma} = -\frac{1}{2} g^{\mu\nu} \frac{\partial I}{\partial x^{\mu}} \frac{\partial I}{\partial x^{\nu}},
\end{eqnarray}
where $I$ represents the Jacobi action for photon motion, and $\sigma$ denotes the affine parameter along the null geodesic. The action can therefore be written in separated form as
\begin{eqnarray}\label{5}
 I = \frac{1}{2} \epsilon^2\sigma - \hat{E} t + \hat{L} \phi + \hat{B}_r(r) + \hat{B}_\theta(\theta),
\end{eqnarray}
where we set $\epsilon=0$, corresponding to the motion of massless particles. The conserved quantities $\hat{E}=pt$ and $\hat{L}=p\phi$ denote the photon energy and the component of angular momentum along the axis of symmetry, respectively. The functions $\hat{B}_r(r)$ and $\hat{B}_\theta(\theta)$ depend exclusively on the radial and angular coordinates, respectively. By substituting the separated form of the action into the Hamilton-Jacobi equation, the equations governing the null geodesic motion of photons are obtained.
\begin{eqnarray}\nonumber
 \rho^2 \frac{dt}{d\sigma} &=& a \big(\hat{L} - a \hat{E} \sin^2\theta \big) + \frac{r^2 + a^2}{\Delta} \big[\hat{E}(r^2 + a^2) - a \hat{L} \big], \\ \nonumber 
 \rho^2 \frac{dr}{d\sigma} &=& \pm \sqrt{\hat{R}(r)},\\ \nonumber
 \rho^2 \frac{d\theta}{d\sigma} &=& \pm \sqrt{\Theta(\theta)}, \\ \label{6}
 \rho^2 \frac{d\phi}{d\sigma} &=& \big(\hat{L} \csc^2\theta - a \hat{E} \big) + \frac{a}{\Delta} [\hat{E}(r^2 + a^2) - a \hat{L}],
\end{eqnarray}
along with
\begin{eqnarray}\nonumber
   \hat{R}(r) &=& [\hat{E}(r^2 + a^2) - a \hat{L}]^2 - \Delta [\hat{J} + (\hat{L} - a \hat{E})^2 ], \\ \label{7}
   \Theta(\theta) &=& \hat{J}+ a^2 \hat{E}^2 - \hat{L}^2 \csc^2\theta \cos^2\theta.
\end{eqnarray}
Here $\hat{J}$ denotes the Carter constant, which arises from the separability of the Hamilton-Jacobi equation and provides an additional conserved quantity for photon motion. The above equations completely determine the null geodesics in the rotating NED spacetime. Of particular interest are photons moving along unstable circular orbits, which define the photon sphere at a radius $r_{ps}$. Such orbits satisfy the conditions for the radial velocity and the radial acceleration to vanish, namely 
$\dot{r}=0$ and $\ddot{r}=0$, where the overdot denotes differentiation with respect to the affine parameter $\sigma$. These requirements are equivalently expressed as 
$\hat{R}(r_{ps})=0$ and $\frac{d\hat{R}}{dr}|_{r=r_{ps}}=0$. Under these conditions, the conserved quantities $\hat{E},~ \hat{L}$, and $\hat{J}$ can be expressed in terms of the corresponding impact parameters, which play a fundamental role in determining the BH shadow.
\begin{eqnarray}\label{8}
\lambda = \frac{\hat{L}}{\hat{E}}, \hspace{1cm} \Lambda  = \frac{\hat{J}}{\hat{E}^2}.  
\end{eqnarray}
Using Eq. (\ref{8}), the impact parameters can be evaluated as follows.
\begin{eqnarray}\label{9}
\lambda(r_{ps})&=& \frac{(a^2+r_{ps}^2) \Delta' (r_{ps})-4 r_{ps} \Delta(r_{ps})}{a \Delta'(r_{ps})},   \\ \label{12} 
\Lambda(r_{ps}) &=& \frac{r_{ps}^2 (-16\Delta(r_{ps})^2 - r_{ps}^2 \Delta'(r_{ps})^2 + 8\Delta(r_{ps})(2a^2 + r_{ps}\Delta'(r_{ps})))}{a^2 \Delta'(r_{ps})^2},
\end{eqnarray}
where the symbol $'$ denotes differentiation with respect to the radial coordinate $r$. To obtain the apparent shadow, we consider a distant observer described by a zero-angular-momentum observer (ZAMO) frame. The observed image is constructed on the observer's screen using the fisheye camera model together with the stereographic projection technique \cite{37,55,56}. The mapping between the four-momentum photon and the celestial coordinates $(\eta,\xi)$ then determines the apparent shape and boundary of the BH shadow \cite{54}.
\begin{eqnarray}\label{10}
\cos\eta= \frac{p^{(1)}}{p^{0}} , \hspace{1cm}  \tan\xi= \frac{p^{(3)}}{p^{(2)}}.
\end{eqnarray}
The observer's image plane is described by a Cartesian coordinate system $(x,y)$, which is related to the corresponding celestial coordinates through an appropriate coordinate transformation.
\begin{eqnarray}\label{11}
  x(r_{ps}) = -2 \tan(\frac{\eta}{2})\sin\xi, \hspace{0.8cm} y(r_{ps}) = -2 \tan(\frac{\eta}{2})\cos\xi.  
\end{eqnarray}
\begin{figure}[H]
    \centering
    \includegraphics[width=0.45\linewidth]{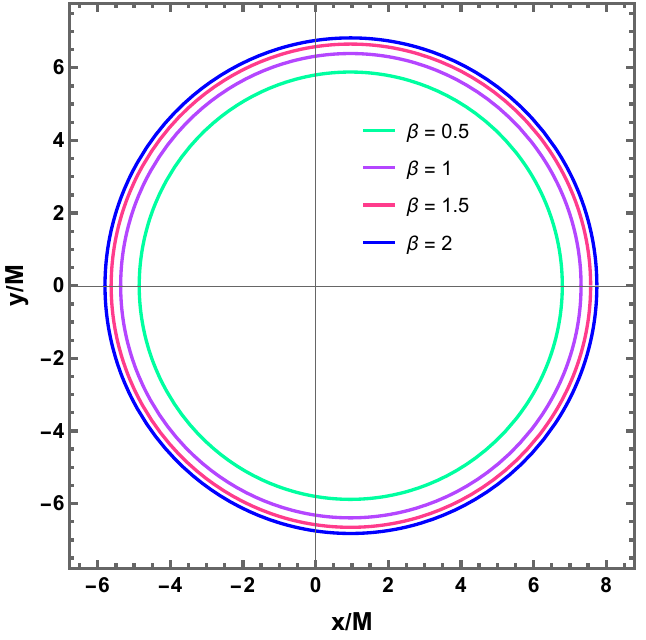}
    \caption{The shadow contours of the rotating NED BH are shown for different values of $\beta$, with fixed $Q=0.5,~a=0.5$, and $\theta_{obs}=80^\circ$.}
    \label{prd2}
\end{figure}
Figure \textbf{\ref{prd2}} illustrates the shadow contours of the rotating NED BH for different values of $\beta$. The shadow is nearly circular for all values of $\beta$, indicating that the NED parameter has only a minor influence on the overall geometry of the shadow. As $\beta$ increases from $0.5$ to $2$, the shadow contour gradually expands outward, resulting in a slight increase in the shadow radius. The corresponding curves remain similar in shape, with only a small separation between them. In general, larger values of $\beta$ produce a comparatively larger shadow. To investigate the geometric properties of the BH shadow, we employ the shadow radius $R_d$ and the distortion parameter $\delta_d$, which are widely used to characterize its size and shape. Throughout this work, these quantities are computed according to the prescription proposed in \cite{57}, which is expressed as
\begin{eqnarray}\label{13}
  R_d= \frac{(x_t-x_r)^2+y_t^2}{2|x_t-x_r|} , \hspace{1cm} \delta_d= \frac{|x_{l'}-x_l|}{2 R_d}.
\end{eqnarray}
To quantify the geometric properties of the BH shadow, a radius reference circle $R_d$ is constructed using the top, bottom, and rightmost points of the shadow boundary. This radius provides a measure of the overall shadow size. The distortion parameter $\delta_d$ is introduced to characterize the departure of the shadow from an ideal circular shape is defined as the horizontal separation between the leftmost edge of the shadow and the corresponding leftmost point of the reference circle. Here, $(x_t, y_t)$, $(x_b, y_b)$, $(x_r, 0)$, and $(x_l, 0)$ denote the top, bottom, rightmost, and leftmost points of the shadow boundary, respectively, whereas $(x_l', 0)$ represents the leftmost point of the reference circle. When the two leftmost points coincide $x_l = x_l'$, the shadow is perfectly circular and the distortion parameter vanishes $\delta_{d}=0$. Any mismatch between these points results in a nonzero value of $\delta_d$, with larger values indicating a greater deviation of the shadow from circular symmetry.

\begin{figure}[H]
\centering
\subfigure[\tiny][~$Q=0.5$]{\label{a1}\includegraphics[width=6.5cm,height=6.5cm]{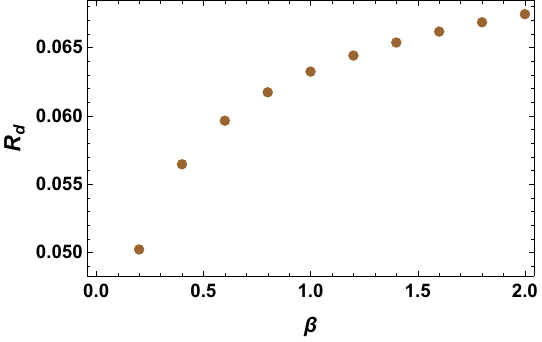}}
\subfigure[\tiny][~$Q=0.5$]{\label{b1}\includegraphics[width=6.5cm,height=6.5cm]{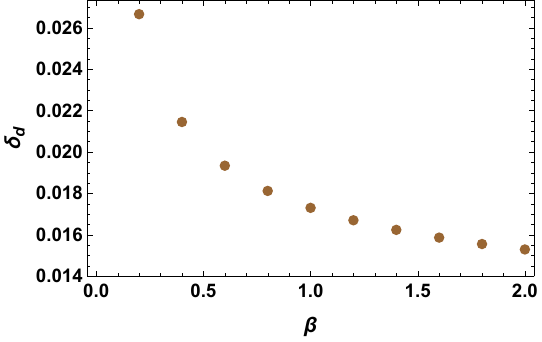}}
\subfigure[\tiny][~$\beta=2$]{\label{c1}\includegraphics[width=6.5cm,height=6.5cm]{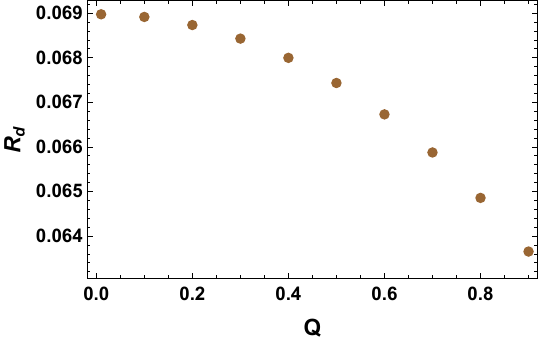}}
\subfigure[\tiny][~$\beta=2$]{\label{d1}\includegraphics[width=6.5cm,height=6.5cm]{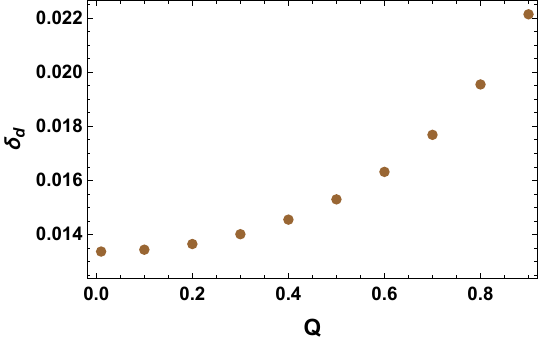}}
\caption{The behavior of $R_d$ and $\delta_d$ for for a fixed value of $Q=0.5$ with varying $\beta$ (upper panels) and for a fixed value of $\beta=2$ with different values of $Q$ (lower panels). In both cases, the observations are made for $\theta_{obs}=80^\circ$, $r_{obs}=100M$, and $a=0.5$.}\label{prd3}
\end{figure}

Figure \textbf{\ref{prd3}} presents the behavior of the shadow observables $R_d$ and $\delta_d$ for different values of  $\beta$ and $Q$. From upper panels, it is shown that the shadow radius $R_d$ increases gradually with increasing $\beta$, whereas the distortion parameter $\delta_d$ exhibits the opposite behavior and decreases monotonically. These results are in agreement with the shadow contours shown in Fig. \textbf{\ref{prd2}}, where the shadow boundary expands while becoming less distorted as $\beta$ increases. The lower panels illustrate the dependence of the shadow observables on $Q$. It can be seen that the shadow radius $R_d$ decreases continuously with increasing $Q$, while the distortion parameter $\delta_d$ increases throughout the considered range. This behavior suggests that a stronger electric charge reduces the apparent size of the BH shadow and simultaneously enhances its deformation from a circular shape. Therefore, the electric charge $Q$ has opposite effects on the shadow size and distortion compared with the NED parameter.

\section{OPTICAL IMAGES IN THE BACKGROUND OF CELESTIAL LIGHT Sphere}
\begin{figure}
\subfigure[\tiny][~$Q=0.1,~\beta=0.5$]{\label{a2}\includegraphics[width=4cm,height=3.8cm]{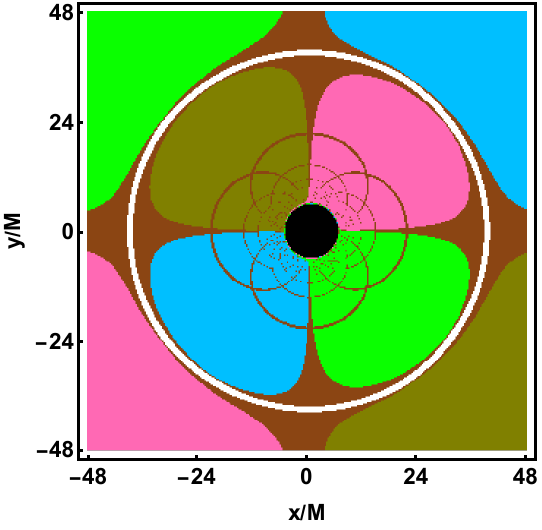}}
\subfigure[\tiny][~$Q=0.1,~\beta=1$]{\label{b2}\includegraphics[width=4cm,height=3.8cm]{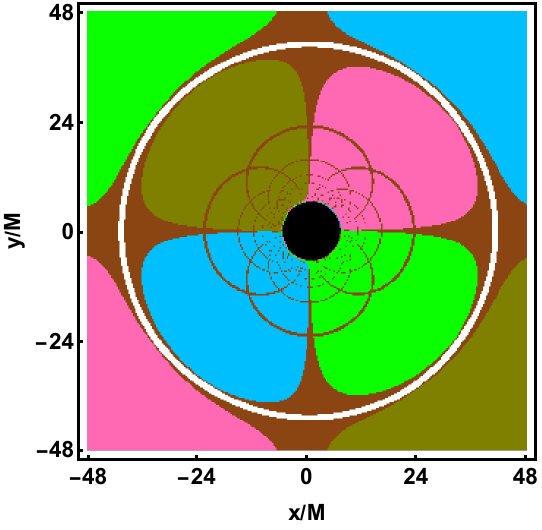}}
\subfigure[\tiny][~$Q=0.1,~\beta=1.5$]{\label{c2}\includegraphics[width=4cm,height=3.8cm]{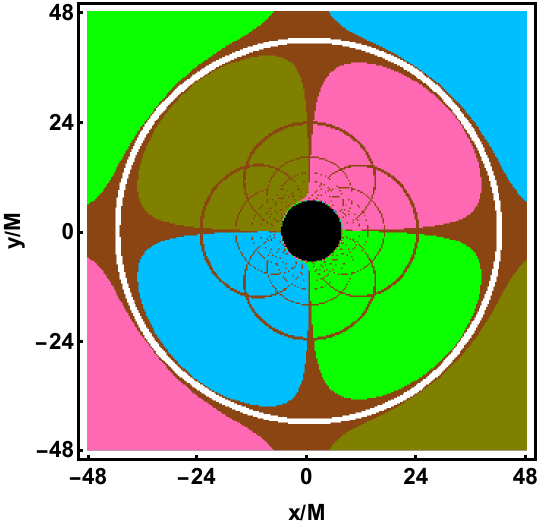}}
\subfigure[\tiny][~$Q=0.1,~\beta=2$]{\label{d2}\includegraphics[width=4cm,height=3.8cm]{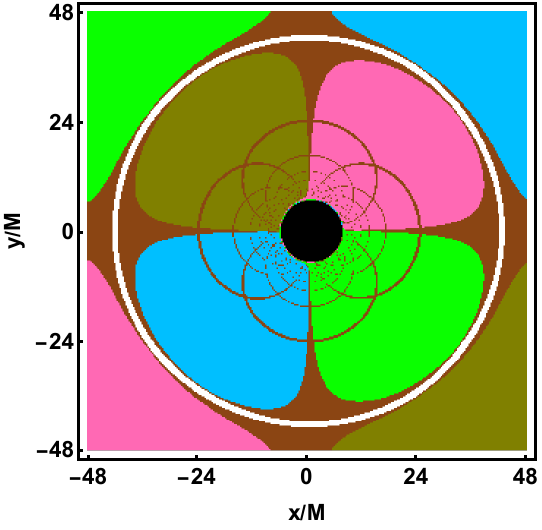}}
\subfigure[\tiny][~$Q=0.5,~\beta=0.5$]{\label{a3}\includegraphics[width=4cm,height=3.8cm]{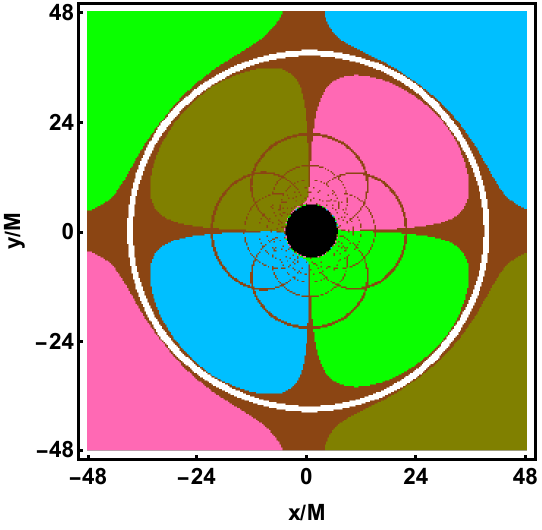}}
\subfigure[\tiny][~$Q=0.5,~\beta=1$]{\label{b3}\includegraphics[width=4cm,height=3.8cm]{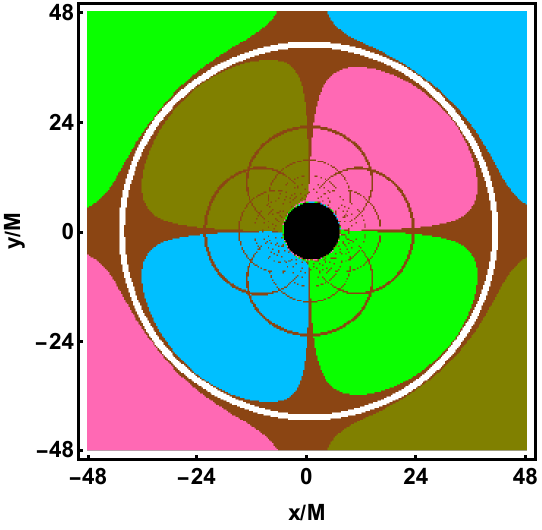}}
\subfigure[\tiny][~$Q=0.5,~\beta=1.5$]{\label{c3}\includegraphics[width=4cm,height=3.8cm]{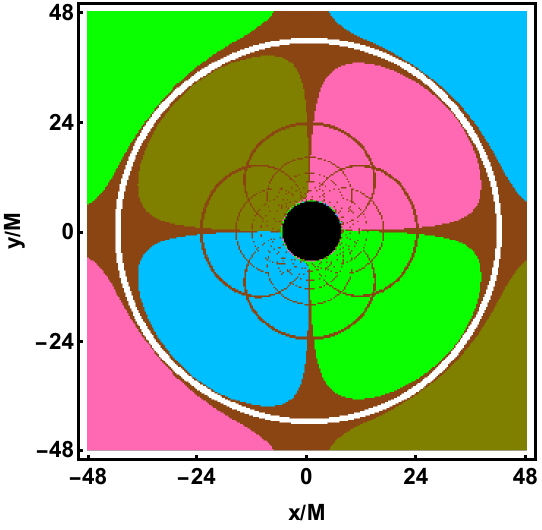}}
\subfigure[\tiny][~$Q=0.5,~\beta=2$]{\label{d3}\includegraphics[width=4cm,height=3.8cm]{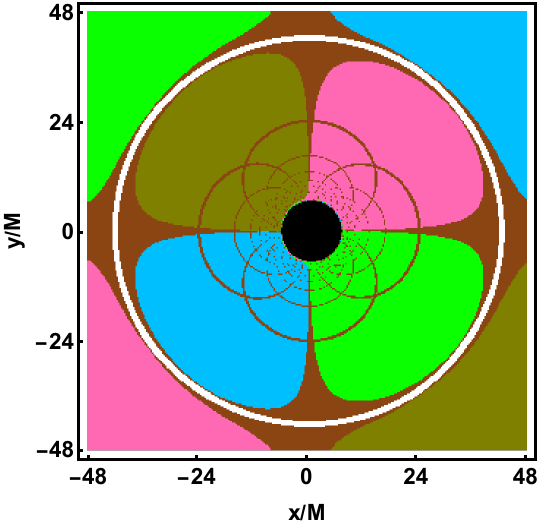}}
\subfigure[\tiny][~$Q=0.9,~\beta=0.5$]{\label{a4}\includegraphics[width=4cm,height=3.8cm]{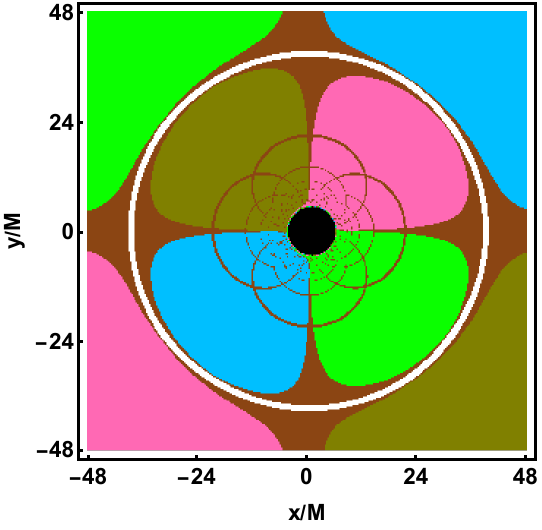}}
\subfigure[\tiny][~$Q=0.9,~\beta=1$]{\label{b4}\includegraphics[width=4cm,height=3.8cm]{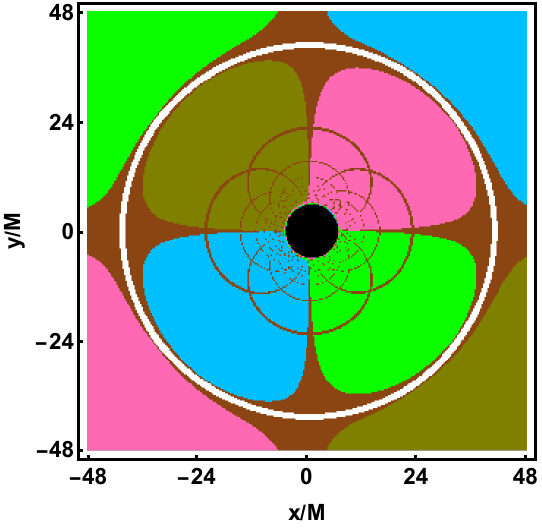}}
\subfigure[\tiny][~$Q=0.9,~\beta=1.5$]{\label{c4}\includegraphics[width=4cm,height=3.8cm]{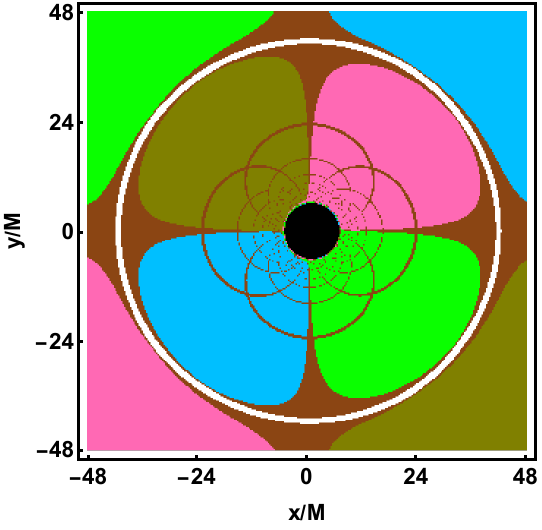}}
\subfigure[\tiny][~$Q=0.9,~\beta=2$]{\label{d4}\includegraphics[width=4cm,height=3.8cm]{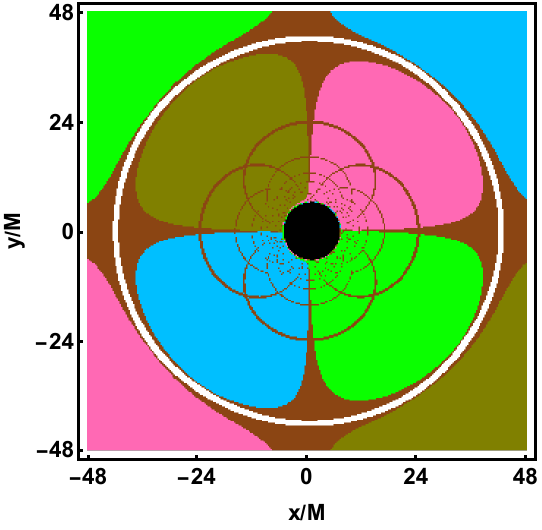}}
\caption{Shadow images of a rotating BH in NED field for different values of $Q$ and $\beta$ with the fixed $a=0.5$, and the observer inclination angle set to $\theta_{obs}=80^\circ$.}\label{prd4}
\end{figure}

To examine the optical appearance of the rotating BH in the presence of a celestial background, we utilize the backward ray-tracing method to reconstruct the observed shadow. In this approach, photons are numerically propagated from the observer's image plane toward the BH, allowing only those trajectories that successfully reach the observer to contribute to the final image. The BH shadow is represented by a dark circular region positioned at the center of a distant celestial sphere, whose radius is taken to be significantly larger than the apparent size of the shadow. To facilitate the interpretation of photon deflection, the celestial sphere is partitioned into four angular regions, each distinguished by a different color (pink, sky blue, olive, and lime). The color distribution in the resulting image provides a clear visualization of the gravitational bending experienced by light rays in the surrounding spacetime. Following the procedure introduced in \cite{55}, the fisheye camera projection is adopted to generate the shadow images. The resulting images are obtained for different values of the NED parameter $\beta$ and the c charge parameter $Q$, while keeping the spin parameter fixed at $a=0.5$ and the observer inclination angle at $\theta_{obs}=80^\circ$, as illustrated in Fig. \textbf{\ref{prd4}}

 Specifically, Fig. \textbf{\ref{prd4}} illustrates that the shadow image of the rotating NED BH under illumination from a celestial light source. The celestial sphere is divided into four angular sectors, shown in pink, sky blue, olive, and lime, to illustrate the deflection of photon trajectories. The longitude and latitude grids are indicated by adjacent reddish-brown lines separated by $10^\circ$. The white circular ring in each panel represents the Einstein ring formed by photons originating from the reference point located along the line joining the BH and the observer. In contrast, the central dark region corresponds to the BH shadow.
 In the first row of Fig. \textbf{\ref{prd4}}, the charge is fixed at $Q=0.1$, while $\beta$ is varied from left to right as $\beta=0.5,~1,~1.5,$ and $2$. It is evident that increasing the parameter $\beta$ gradually enlarges the apparent shadow of BH. The characteristic ``D''-shaped petals enclosed by the Einstein ring remain almost unchanged throughout the variation of $\beta$, indicating that their morphology is only weakly influenced by the NED effects. In contrast, the Einstein ring gradually expands as $\beta$ increases, becoming more pronounced while preserving its nearly circular shape. These results suggest that the NED parameter primarily affects the overall size of the shadow and the Einstein ring without producing any significant modification in the internal petal-like structure. The second and third rows illustrate the corresponding shadow images for larger charge values, namely $Q=0.5$ and $Q=0.9$, respectively, while the same sequence of $\beta$ is retained. For each fixed value of $Q$, the influence of the NED parameter remains qualitatively the same. As $\beta$ increases, the shadow radius grows, the distortion decreases, the ``D''-shaped petals preserve their overall structure, and the Einstein ring continues to expand slightly. These results indicate that the NED parameter primarily governs the apparent size of the shadow and the Einstein ring, while exerting only a minor influence on the internal petal-like structure.

A comparison of the three rows reveals that the charge parameter $Q$ produces an effect opposite to that of $\beta$. As $Q$ increases from $0.1$ to $0.5$ and finally to $0.9$, the apparent size of the BH shadow gradually decreases, whereas the distortion parameter increases, making the shadow boundary more asymmetric. The Einstein ring also becomes slightly smaller with increasing $Q$, while the ``D''-shaped petals exhibit a more pronounced deformation compared to the lower $Q$ cases. Therefore, unlike the NED parameter, which enlarges the shadow and suppresses its distortion, the charge parameter reduces the shadow size and enhances its asymmetry. This opposite behavior demonstrates the competing roles of $\beta$ and $Q$ in shaping the observable appearance of the BH shadow.
\section{Black Hole Imaging with Thin Accretion Disk Emission}
The investigation of thin accretion disks around BHs provides an effective framework for understanding the interaction between strong gravitational fields and the surrounding radiating matter. In this section, we examine the properties of thin accretion disks in the spacetime of a rotating BH arising from NED. Our primary objective is to explore how the BH spin, electric charge, and NED parameters modify the dynamics of the accreting matter and influence the resulting observational signatures. Throughout the analysis, the accreting material is modeled as electrically neutral particles moving along equatorial timelike circular geodesics, which is consistent with the standard relativistic thin disk description. 
An important quantity governing the accretion process is the ISCO, which determines the inner boundary of the accretion disk. The location of the ISCO plays a central role in the disk's emission properties, as it controls the efficiency with which the rest-mass energy of infalling matter is converted into radiation. Particles located outside the ISCO remain on stable circular orbits and continuously contribute to the disk luminosity. In contrast, those crossing this radius lose orbital stability and rapidly plunge toward the BH. Consequently, the determination of the ISCO provides essential information for understanding the structure of the accretion disk and the observable optical appearance of the rotating NED BH \cite{40}. In this regard, the ISCO position of a
BH is given by
\begin{eqnarray}\label{14}
    V_{eff}(r)=0, \hspace{1cm} \partial_r V_{eff}=0,
\end{eqnarray}
where $V_{eff}$ denotes the effective potential, which is given by
\begin{eqnarray}\label{15}
\mathcal{V}_{eff}=(1+g^{tt}{\mathcal{E}}+g^{\phi \phi}{\mathcal{L}}-2g^{t\phi} {\mathcal{E}}{\mathcal{L}}).
\end{eqnarray}
The conserved energy and angular momentum of the particle can be written as
\begin{eqnarray}\label{16}
{\mathcal{E}}=-\frac{1}{\sqrt{f_{1}}}(g_{tt}+g_{t\phi}{\Omega}), \quad {\mathcal{L}}=\frac{1}{\sqrt{f_{1}}}(g_{t\phi}+g_{\phi\phi}{\Omega}),
\end{eqnarray}
here
\begin{eqnarray}\label{17a}
f_{1}=-g_{tt}-2g_{t\phi}{\Omega}-g_{\phi\phi}{\Omega}^{2},  \textit{and} \quad
\Omega=\frac{d\phi}{dt}=\frac{\partial_{r}g_{t\phi}+\sqrt{(\partial_{r}g_{t\phi})^{2}-\partial_{r}g_{tt}\,\partial_{r}g_{\phi\phi}}}{\partial_{r}g_{\phi\phi}}.
\end{eqnarray}
 At the ISCO radius, $r=r_{ISCO}$ the corresponding conserved quantities are represented by $\mathcal{E}_{ISCO}$ and $\mathcal{L}_{ISCO}$. For radii satisfying $r>r_{ISCO}$, the accreting matter moves along stable Keplerian circular orbits, and its four velocity is given by
\begin{eqnarray}\label{17}
v^{\alpha}_{out}=\frac{1}{\sqrt{f_{1}}}(1,0,0,{\Omega}).
\end{eqnarray}
Within ISCO $r<r_{ISCO}$, the accreting matter leaves its stable circular orbit and plunges toward the event horizon along the critical particles falling. During this phase, the particles retain the conserved energy and angular momentum acquired at the ISCO. The corresponding components of the four-velocity are therefore given by \cite{40}
\begin{eqnarray}\nonumber
v^{t}_{plung}&=&(-g^{tt}{\hat{E}}_{ISCO}+g^{t\phi}{\hat{L}}_{ISCO}),\quad
v^{\phi}_{plung}=(-g^{t\phi}{\hat{E}}_{ISCO}+g^{\phi\phi}{\hat{L}}_{ISCO}),\\\nonumber
v^{r}_{plung}&=&-\big(-(g_{tt}v^{t}_{plung}v^{t}_{plung}+2g_{t\phi}v^{t}_{plung}
v^{\phi}_{plung}+g_{\phi\varphi}v^{\phi}_{plung}v^{\phi}_{plung}+1)(g_{rr})^{-1}\big)^{\frac{1}{2}},\\\label{18}~v^{\theta}_{plung}&=&0.
\end{eqnarray}
Photons emitted from the accretion disk may intersect the disk plane one or more times before reaching a distant observer. Depending on the number of intersections, the resulting images are classified as direct $n=1$, lensed $n=2$, and higher-order images $n>2$. In the present study, we focus exclusively on the direct and lensed images, as they provide the dominant contribution to the observed appearance of the BH. As photons propagate through the disk, their observed intensity is primarily determined by the local emission and absorption processes, while the effects of reflection are neglected for simplicity. Under these assumptions, the specific intensity measured on the observer's image plane is expressed as
\begin{eqnarray}\label{19}
\mathcal{S}_{obs}=\sum_{n=1}^{N_{max}}f_{n}j_{n}^{3}(r_{n})K_{n}.    
\end{eqnarray}
Here, $\mathcal{S}_{obs}$ denotes the photon frequency measured on the observer's image plane, while $N_max$ specifies the maximum number of times that a photon intersects the accretion disk before reaching the observer. The quantity $j_n$ represents the corresponding redshift factor, and $K_n$ denotes the emissivity evaluated at the $ n^ {th} $ intersection $n^th$ with the disk. Throughout this work, the normalization factor is taken to be $f_n=1$. The emissivity function $K_n$ is expressed as
\begin{equation}\label{24}
\tau_{n}=\exp\big[\tau_{1}g^{2}+\tau_{2}g\big],
\end{equation}
Here, $g=\log(r/r_+)$, while the constants are chosen as $\tau_1=-1/2$ and $\tau_2=-2$ \cite{58}. These parameter values are consistent with the horizon-scale images of BHs obtained by the Event Horizon Telescope at a wavelength of $1.3$ mm ($230$ GHz) \cite{58}. It should be noted that the redshift factor exhibits distinct behavior in the regions outside and inside the ISCO, reflecting the different dynamical properties of the emitting matter. Outside the ISCO, particles move on stable Keplerian circular orbits, whereas inside the ISCO they plunge toward the event horizon, leading to different redshift characteristics. The expression for the redshift factor in the region outside the ISCO is given by \cite{59}
\begin{equation}\label{21}
j^{out}_{n}=\frac{\omega_0(1-\psi\frac{p_{\phi}}{p_{t}})}{\sigma_0(1+{\Psi}\frac{p_{\phi}}{p_{t}})}|_{r=r_{n}},
\quad \quad r\geq r_{ISCO},
\end{equation}
where 
$\omega_0=\sqrt{\frac{g_{\phi\phi}}{g^2_{t\phi}-g_{tt}g_{\phi\phi}}}$,~$\lambda_0=\frac{g_{t\phi}}{g_{\phi\phi}}$,~
$\sigma_o=\frac{1}{\sqrt{f_{1}}}$ and
$\hat{e}=\frac{p_{(t)}}{p_{t}}=\omega_0(1-\psi_0\frac{p_{\phi}}{p_{t}})$. The connection between the photon energy measured by a distant observer and the energy carried along a null geodesic is described by the redshift factor. For an asymptotically flat spacetime, a static observer located at spatial infinity measures the normalized energy as $\hat{e}=1$. In contrast, within the region $r<r_ISCO$, the accreting matter no longer follows stable circular orbits but instead plunges toward the event horizon along the critical trajectory. As a result, the corresponding redshift factor takes a different form, which is given by \cite{59}
\begin{eqnarray}\label{22}
j^{plung}_{n}=-\frac{1}{u^{r}_{plung}p_{r}/p_{t}-{\mathcal{E}}_{ISCO}(g^{tt}-g^{t\phi}p_{\phi}/p_{t})
+{\mathcal{L}}_{ISCO}(g^{\phi\phi}p_{\phi}/p_{t}+g^{t\phi})}|_{r=r_{n}},
 \quad r< r_{ISCO}.
\end{eqnarray}
\begin{figure}
    \centering
    \includegraphics[width=0.45\linewidth]{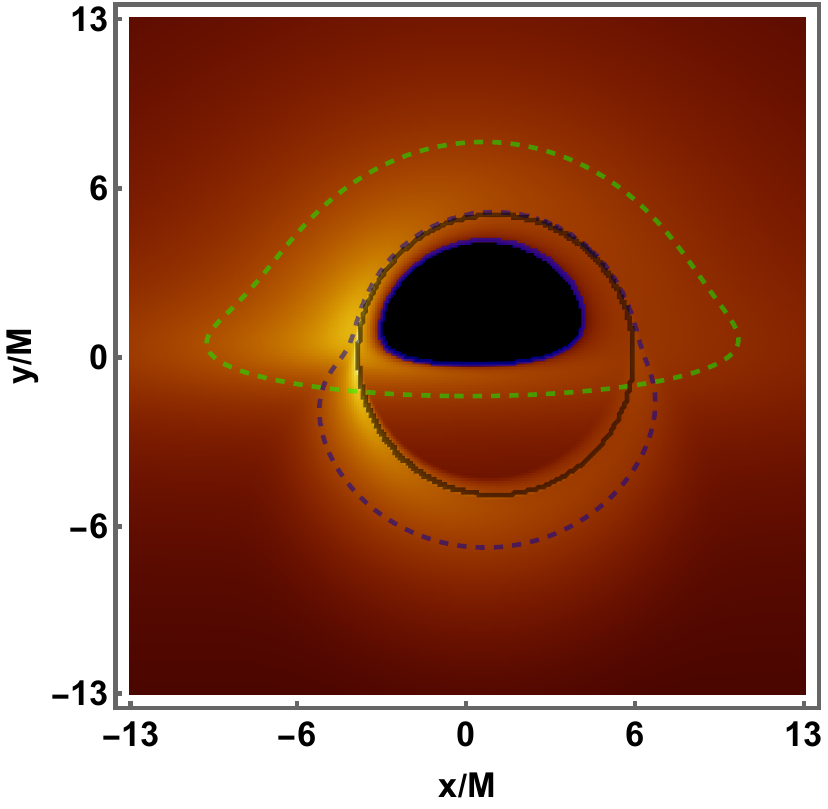}
    \caption{Physical interpretation of the four characteristic curves for the rotating BH in NED field with $a=0.5,~ Q=0.7,~\beta=0.9$, and $\theta_{obs}=80^\circ$ under prograde accretion flow. The green dashed and blue solid curves represent the direct images of the accretion disk at $r=9$ and $r=r_+$, respectively. The blue dashed curve shows the lensed image of the accretion disk at $r=9$, whereas the black solid curve represents the critical curve of the rotating BH.}
    \label{NED}
\end{figure}

\begin{figure}
\subfigure[\tiny][~$Q=0.1,~\beta=0.5$]{\label{a5}\includegraphics[width=4cm,height=3.5cm]{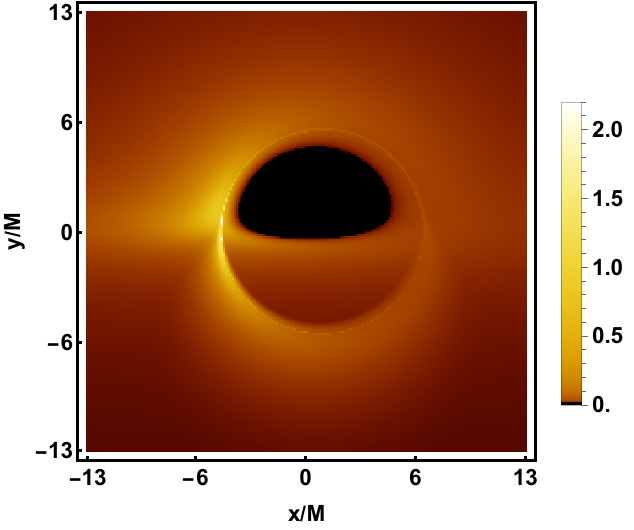}}
\subfigure[\tiny][~$Q=0.1,~\beta=1$]{\label{b5}\includegraphics[width=4cm,height=3.5cm]{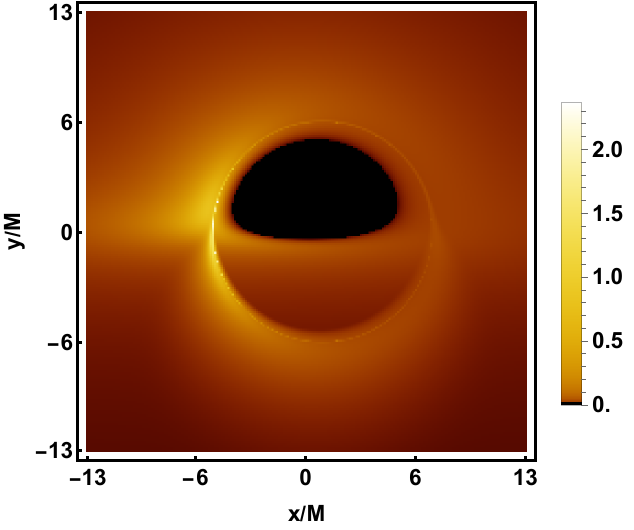}}
\subfigure[\tiny][~$Q=0.1,~\beta=1.5$]{\label{c5}\includegraphics[width=4cm,height=3.5cm]{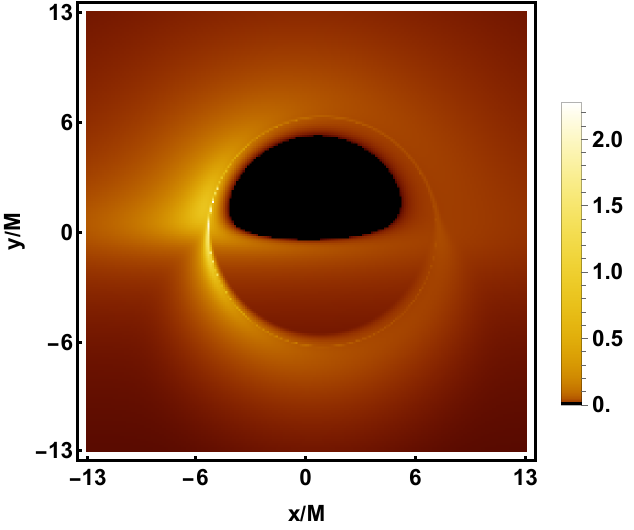}}
\subfigure[\tiny][~$Q=0.1,~\beta=2$]{\label{d5}\includegraphics[width=4cm,height=3.5cm]{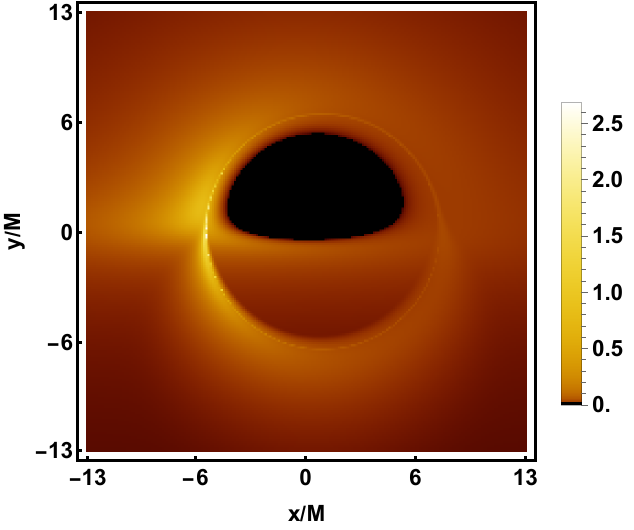}}
\subfigure[\tiny][~$Q=0.5,~\beta=0.5$]
{\label{a6}\includegraphics[width=4cm,height=3.5cm]{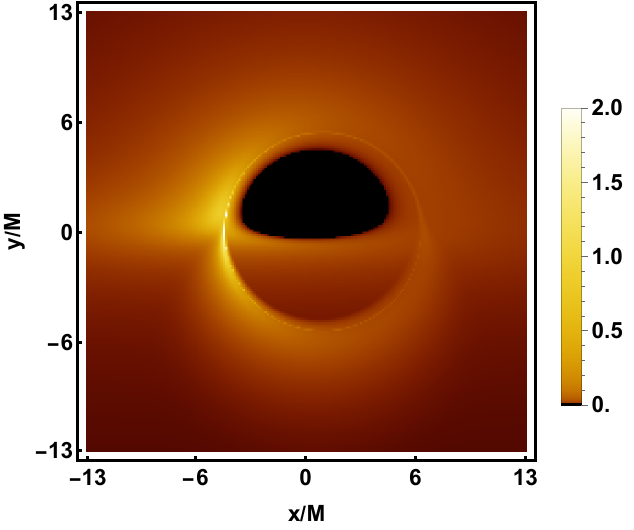}}
\subfigure[\tiny][~$Q=0.5,~\beta=1$]{\label{b6}\includegraphics[width=4cm,height=3.5cm]{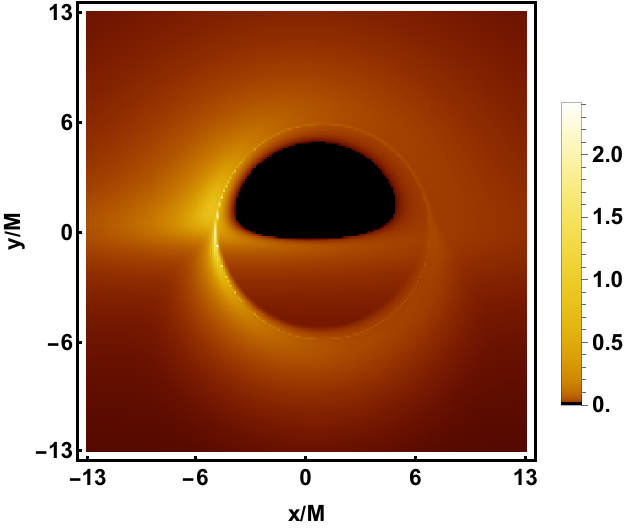}}
\subfigure[\tiny][~$Q=0.5,~\beta=1.5$]{\label{c6}\includegraphics[width=4cm,height=3.5cm]{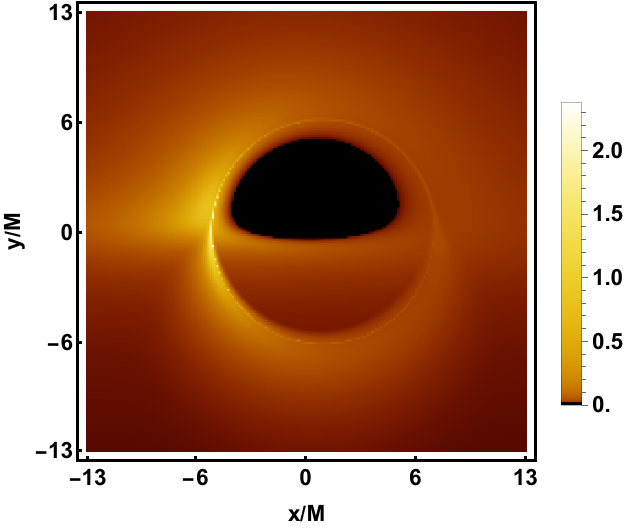}}
\subfigure[\tiny][~$Q=0.5,~\beta=2$]{\label{d6}\includegraphics[width=4cm,height=3.5cm]{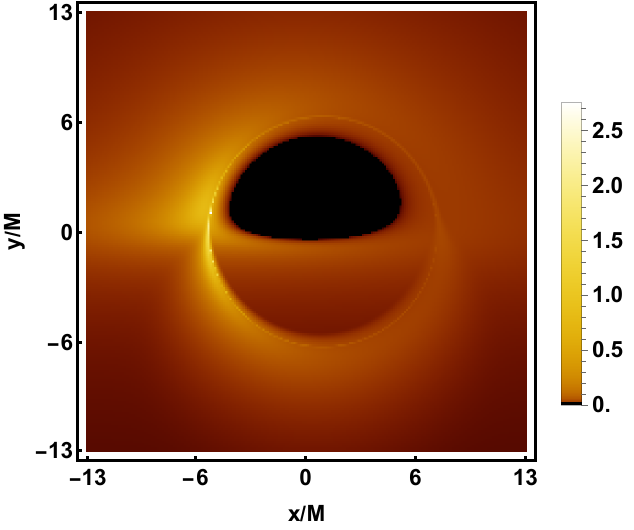}}
\subfigure[\tiny][~$Q=0.9,~\beta=0.5$]
{\label{a7}\includegraphics[width=4cm,height=3.5cm]{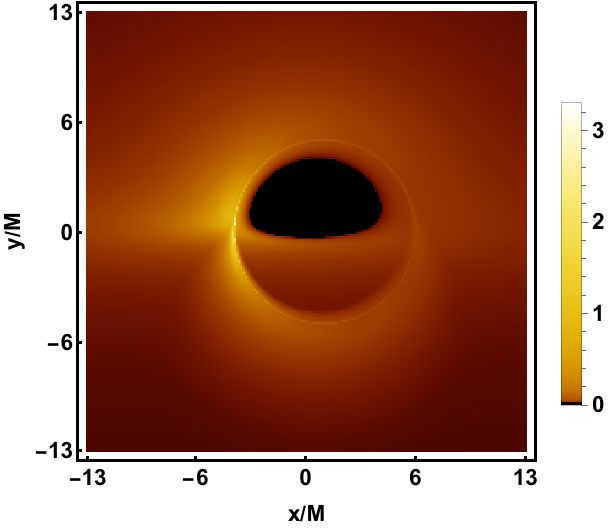}}
\subfigure[\tiny][~$Q=0.9,~\beta=1$]{\label{b7}\includegraphics[width=4cm,height=3.5cm]{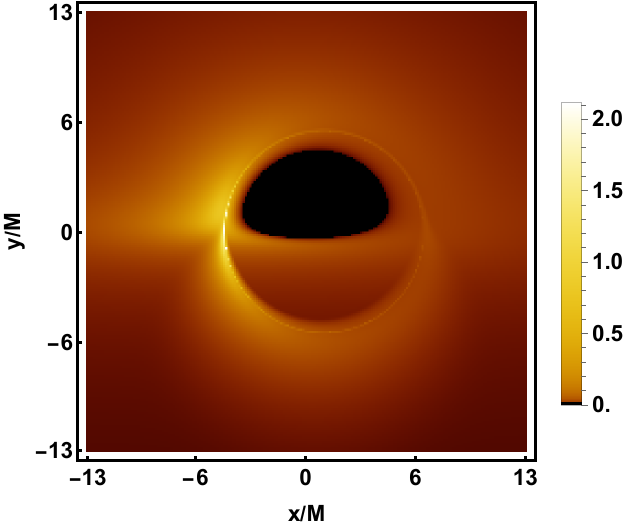}}
\subfigure[\tiny][~$Q=0.9,~\beta=1.5$]{\label{c7}\includegraphics[width=4cm,height=3.5cm]{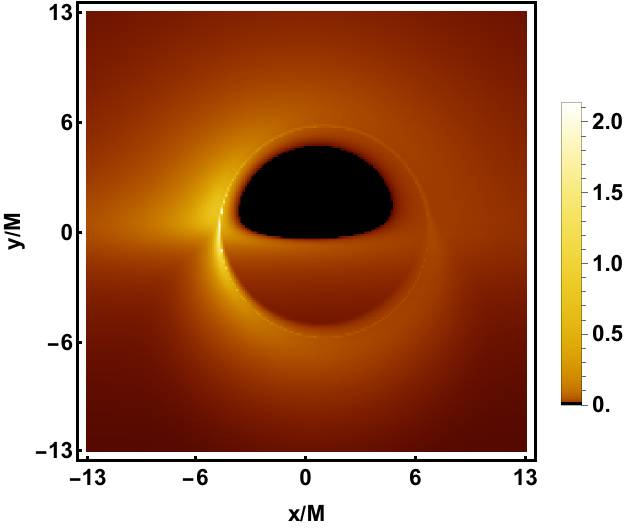}}
\subfigure[\tiny][~$Q=0.9,~\beta=2$]{\label{d7}\includegraphics[width=4cm,height=3.5cm]{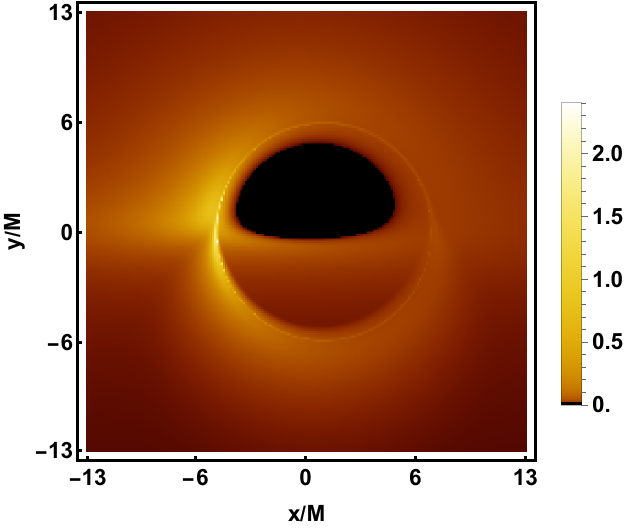}}
\caption{Optical images of the rotating BH in NED field under prograde accretion flow for different values of $Q$ and $\beta$. The observer inclination is fixed at $\theta_{obs} = 80^\circ$, and $a=0.5$.}\label{prd5}
\end{figure}

Figure~\textbf{\ref{NED}} presents the four characteristic curves of the rotating BH in NED field. The green dashed and blue solid curves correspond to the direct images of the accretion disk at $r=9$ and $r=r_+$, respectively, with the latter representing the inner shadow of the BH \cite{58}. The blue dashed curve shows the lensed image of the accretion disk at $r=9$, whereas the black solid curve denotes the critical curve, which is identified as the photon ring. Together, these curves describe the main optical features of the rotating NED BH and provide the basis for understanding the formation of the observed shadow and accretion-disk images.
In Fig. \textbf{\ref{prd5}} presents the optical appearance of the rotating NED BH surrounded by a thin accretion disk for different values of the NED and electric charge parameters. The images are obtained for a fixed observer inclination angle $\theta_{obs} = 80^\circ$, and $a=0.5$. In all panels, a bright photon ring enclosing a dark central region is clearly visible. The bright ring is produced by photons that experience strong gravitational lensing before reaching the observer, whereas the central dark region corresponds to the BH shadow. From left to right, the value of $\beta$ increases from $0.5$ to $2$ while the electric charge remains fixed in each row. As $\beta$ increases, the shadow boundary gradually expands, leading to a larger shadow radius. At the same time, the shadow becomes more symmetric, and the distortion of its boundary is slightly reduced. The influence of the electric charge is illustrated from the top to the bottom, corresponding to $Q=0.1,~0.5$, and $0.9$, respectively. As the charge increases, the apparent size of the shadow decreases, while the deformation of the shadow boundary becomes more noticeable. Consequently, the photon ring appears closer to the central dark region, producing a more compact image. These results are in agreement with Fig. \textbf{\ref{prd3}}.

\begin{figure}
\subfigure[\tiny][~$Q=0.1,~\beta=0.5$]{\label{a8}\includegraphics[width=4cm,height=3.5cm]{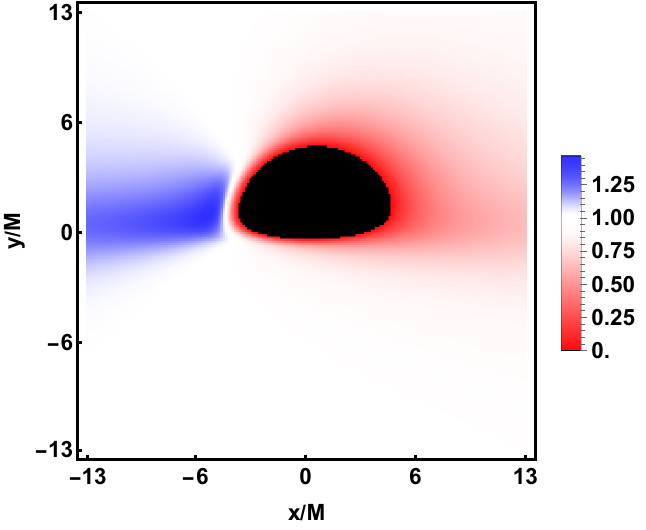}}
\subfigure[\tiny][~$Q=0.1,~\beta=1$]{\label{b8}\includegraphics[width=4cm,height=3.5cm]{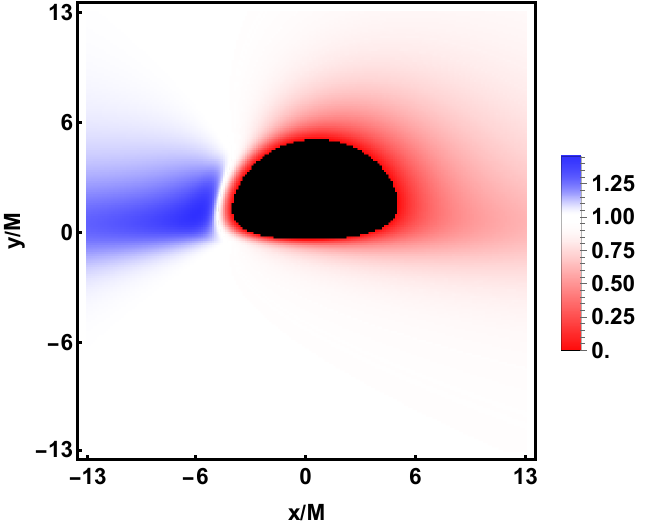}}
\subfigure[\tiny][~$Q=0.1,~\beta=1.5$]{\label{c8}\includegraphics[width=4cm,height=3.5cm]{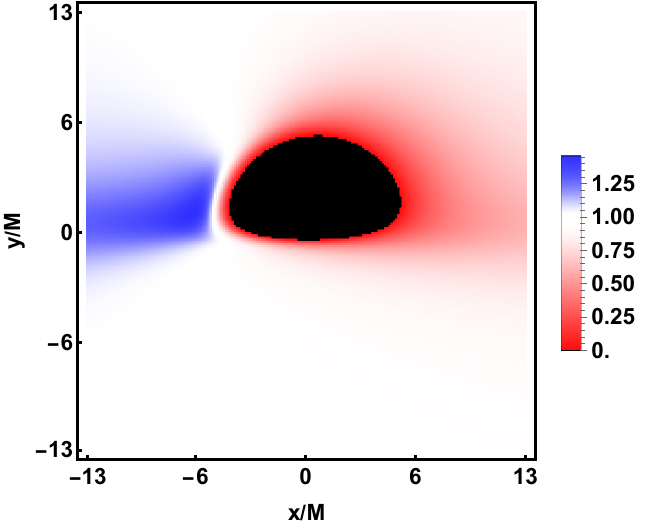}}
\subfigure[\tiny][~$Q=0.1,~\beta=2$]{\label{d8}\includegraphics[width=4cm,height=3.5cm]{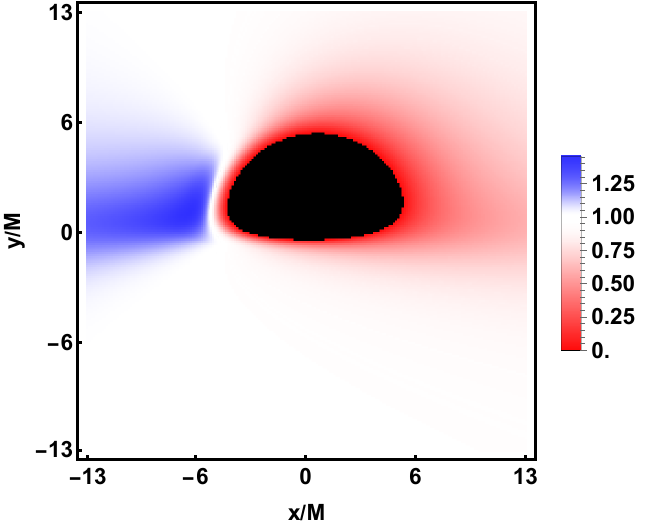}}
\subfigure[\tiny][~$Q=0.5,~\beta=0.5$]
{\label{a9}\includegraphics[width=4cm,height=3.5cm]{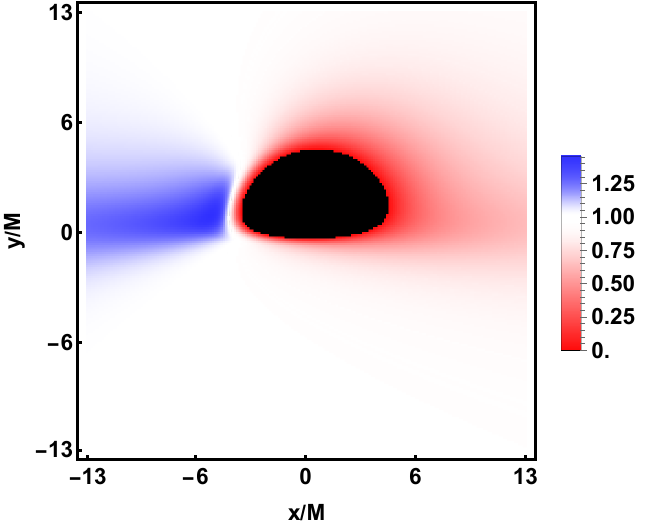}}
\subfigure[\tiny][~$Q=0.5,~\beta=1$]{\label{b9}\includegraphics[width=4cm,height=3.5cm]{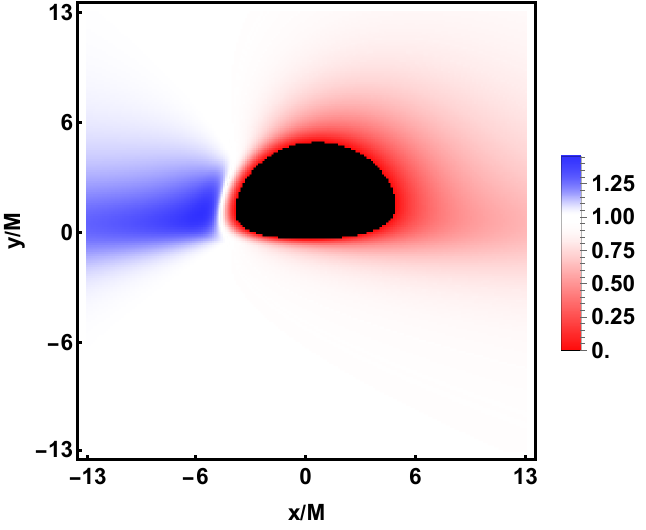}}
\subfigure[\tiny][~$Q=0.5,~\beta=1.5$]{\label{c9}\includegraphics[width=4cm,height=3.5cm]{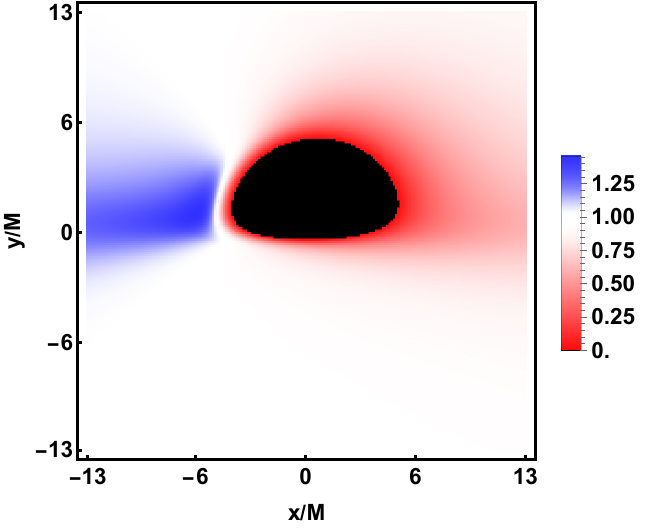}}
\subfigure[\tiny][~$Q=0.5,~\beta=2$]{\label{d9}\includegraphics[width=4cm,height=3.5cm]{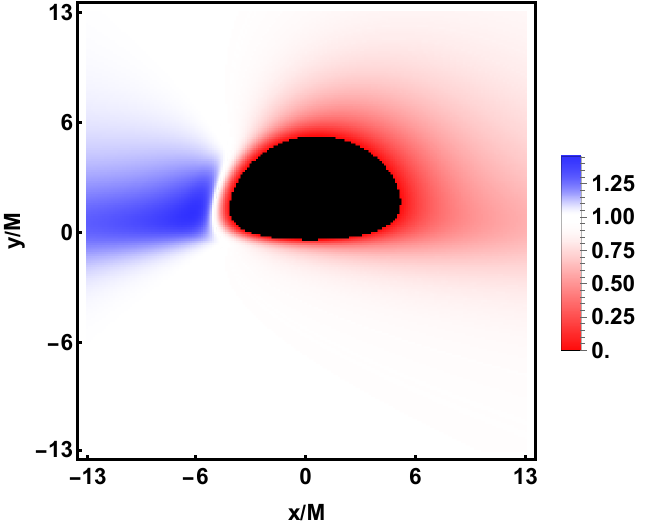}}
\subfigure[\tiny][~$Q=0.9,~\beta=0.5$]
{\label{a10}\includegraphics[width=4cm,height=3.5cm]{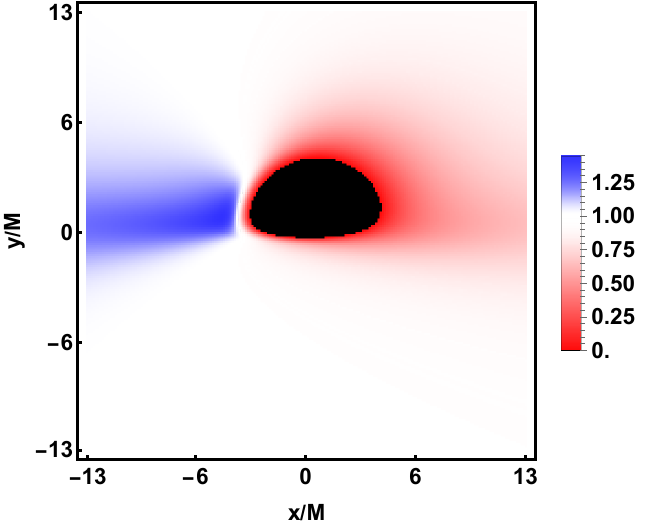}}
\subfigure[\tiny][~$Q=0.9,~\beta=1$]{\label{b10}\includegraphics[width=4cm,height=3.5cm]{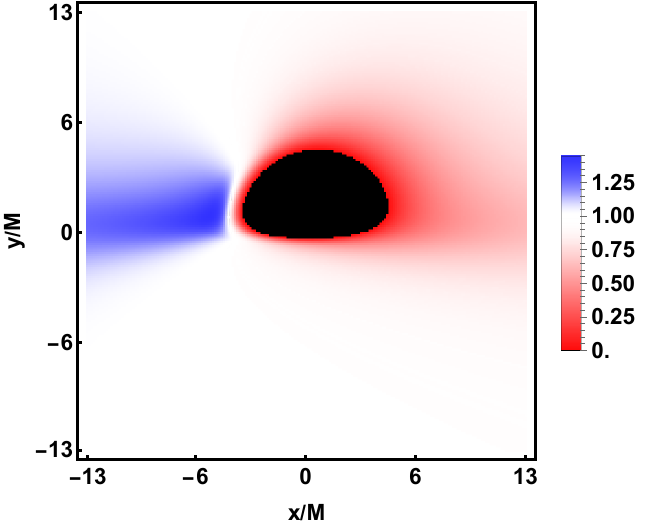}}
\subfigure[\tiny][~$Q=0.9,~\beta=1.5$]{\label{c10}\includegraphics[width=4cm,height=3.5cm]{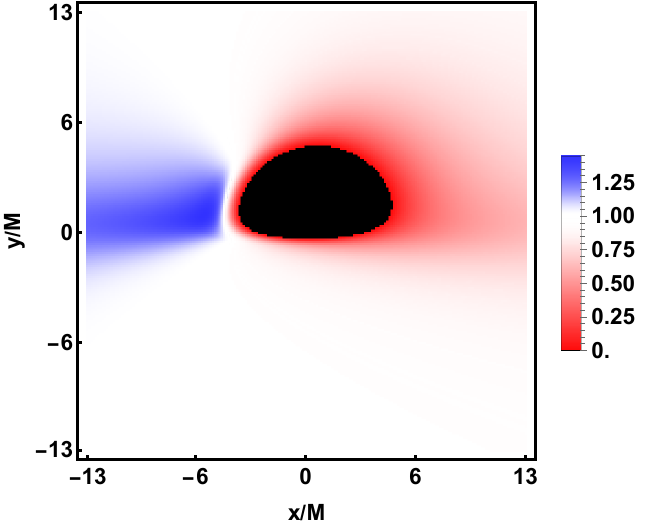}}
\subfigure[\tiny][~$Q=0.9,~\beta=2$]{\label{d10}\includegraphics[width=4cm,height=3.5cm]{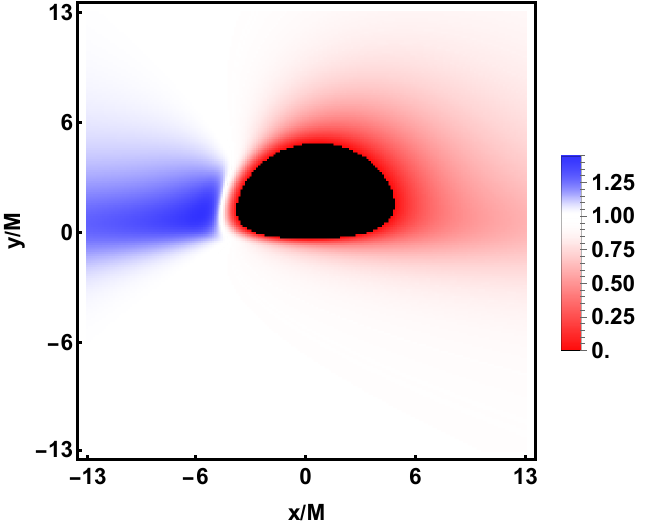}}
\caption{The redshift distribution of direct images for the rotating BH in NED field is presented for different values of $Q$ and $\beta$, with the observer's inclination fixed at $\theta_{obs}=80^\circ$, and $a=0.5$ under prograde accretion flow.}\label{prd6}
\end{figure}
In Fig. \textbf{\ref{prd6}}, we illustrates the redshift distribution of the direct images for the rotating NED BH under the same set of parameters adopted in Fig. \textbf{\ref{prd5}}. The blue and red colors represent the blueshifted and redshifted emission, respectively, while the central dark region corresponds to the BH shadow. From left to right, the values of $\beta$ increases from $0.5$ to $2$ for a fixed value of $Q$. As $\beta$ increases, the redshifted region gradually spreads over a larger portion of the image, particularly around the photon ring. In contrast, the blueshifted region also expands slightly while preserving its location on the opposite side of the disk. The influence of the electric charge is shown from the top to the bottom rows, corresponding to $Q=0.1,~ 0.5$, and $0.9$, respectively. As $Q$ increases, the redshifted and blueshifted regions become more compact and move closer to the central shadow due to the reduction in the shadow radius. In addition, the asymmetry between the two regions becomes more pronounced, reflecting the enhanced distortion of the shadow boundary at larger values of $Q$.
\begin{figure}
\subfigure[\tiny][~$Q=0.1,~\beta=0.5$]{\label{a11}\includegraphics[width=4cm,height=3.5cm]{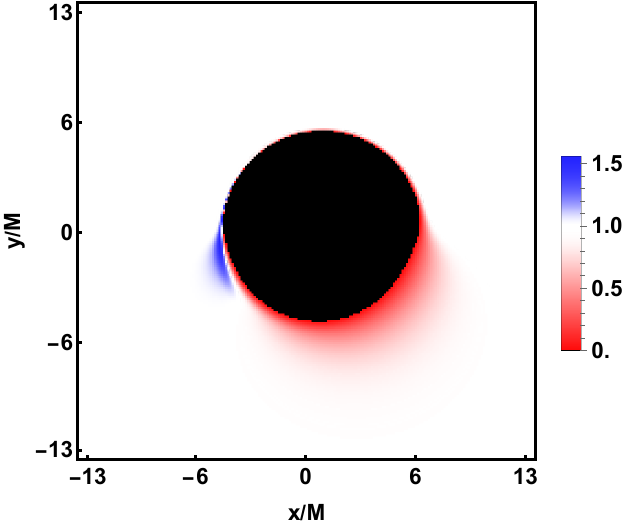}}
\subfigure[\tiny][~$Q=0.1,~\beta=1$]{\label{b11}\includegraphics[width=4cm,height=3.5cm]{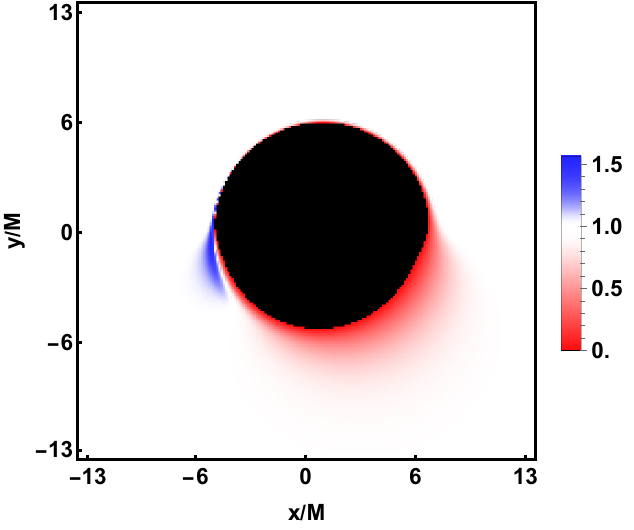}}
\subfigure[\tiny][~$Q=0.1,~\beta=1.5$]{\label{c11}\includegraphics[width=4cm,height=3.5cm]{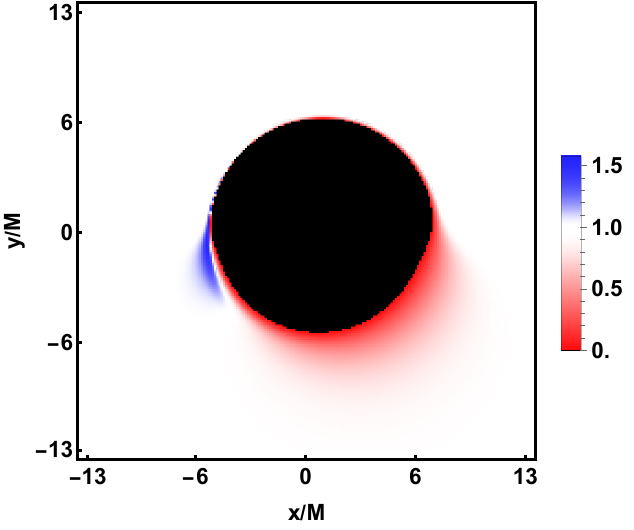}}
\subfigure[\tiny][~$Q=0.1,~\beta=2$]{\label{d11}\includegraphics[width=4cm,height=3.5cm]{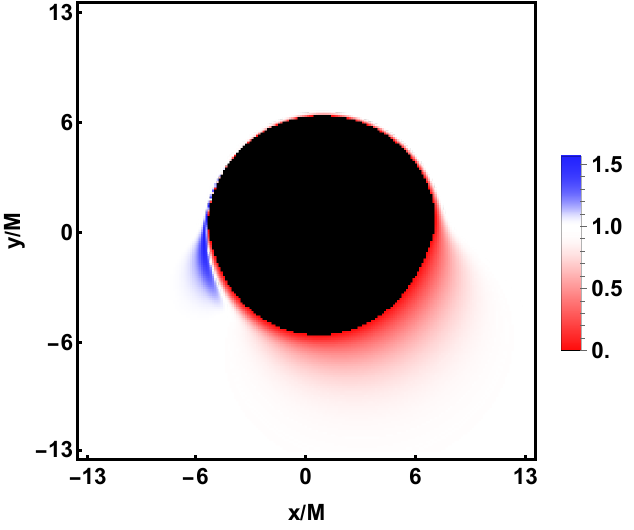}}
\subfigure[\tiny][~$Q=0.5,~\beta=0.5$]
{\label{a12}\includegraphics[width=4cm,height=3.5cm]{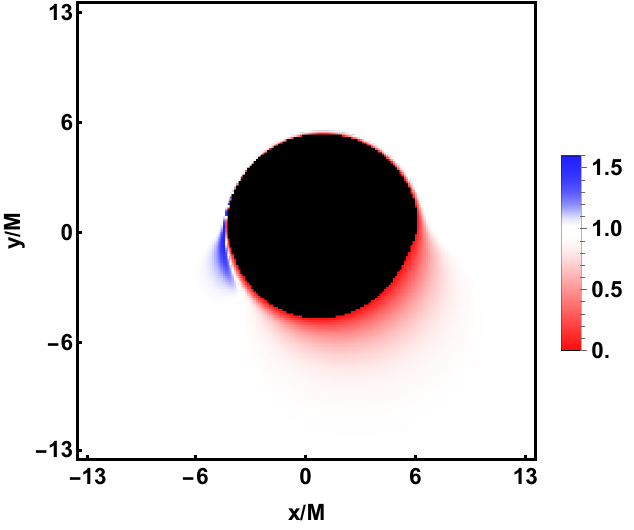}}
\subfigure[\tiny][~$Q=0.5,~\beta=1$]{\label{b12}\includegraphics[width=4cm,height=3.5cm]{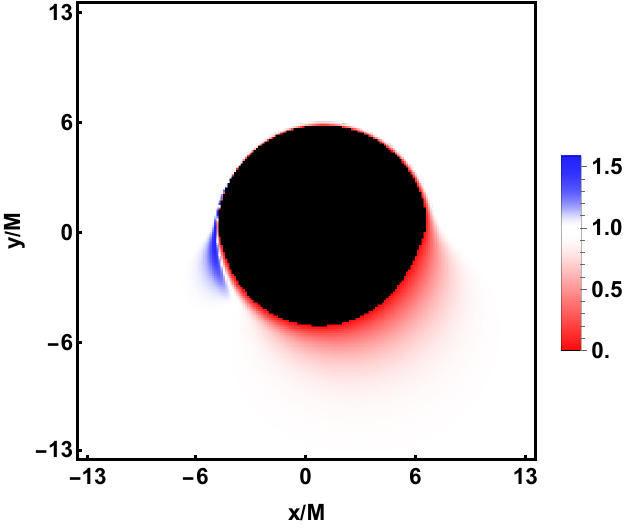}}
\subfigure[\tiny][~$Q=0.5,~\beta=1.5$]{\label{c12}\includegraphics[width=4cm,height=3.5cm]{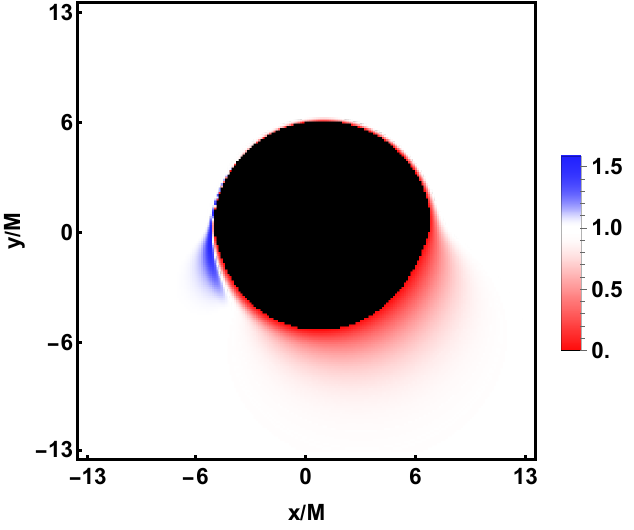}}
\subfigure[\tiny][~$Q=0.5,~\beta=2$]{\label{d12}\includegraphics[width=4cm,height=3.5cm]{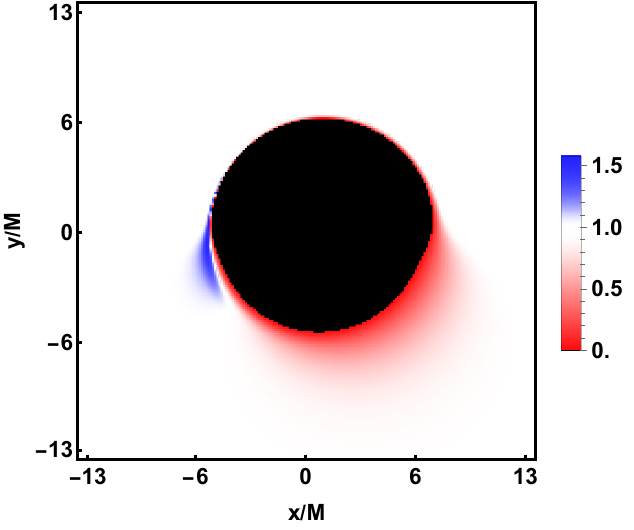}}
\subfigure[\tiny][~$Q=0.9,~\beta=0.5$]
{\label{a13}\includegraphics[width=4cm,height=3.5cm]{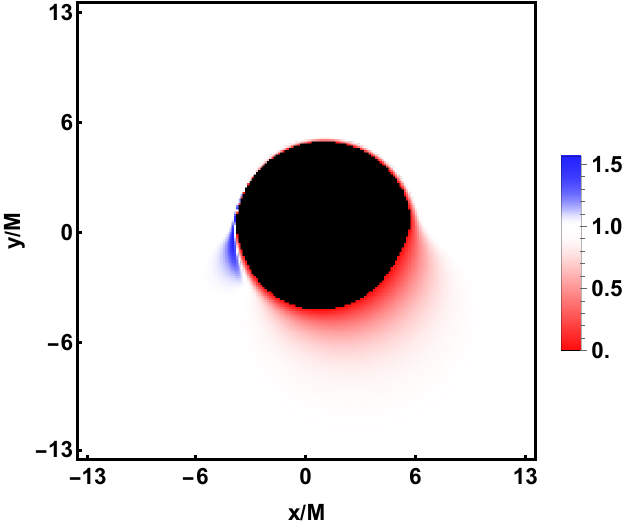}}
\subfigure[\tiny][~$Q=0.9,~\beta=1$]{\label{b13}\includegraphics[width=4cm,height=3.5cm]{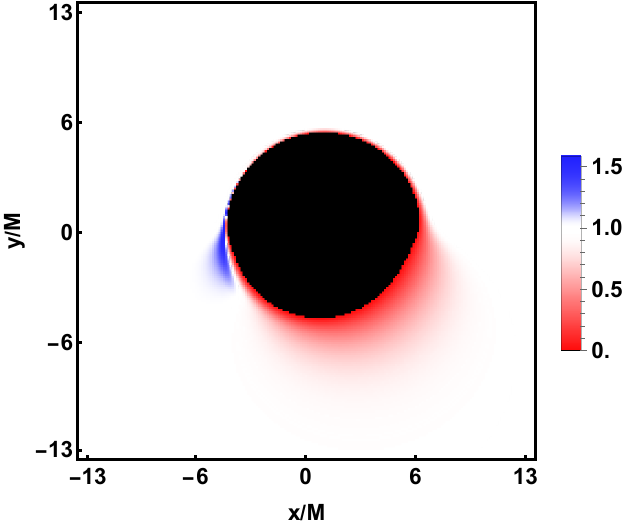}}
\subfigure[\tiny][~$Q=0.9,~\beta=1.5$]{\label{c13}\includegraphics[width=4cm,height=3.5cm]{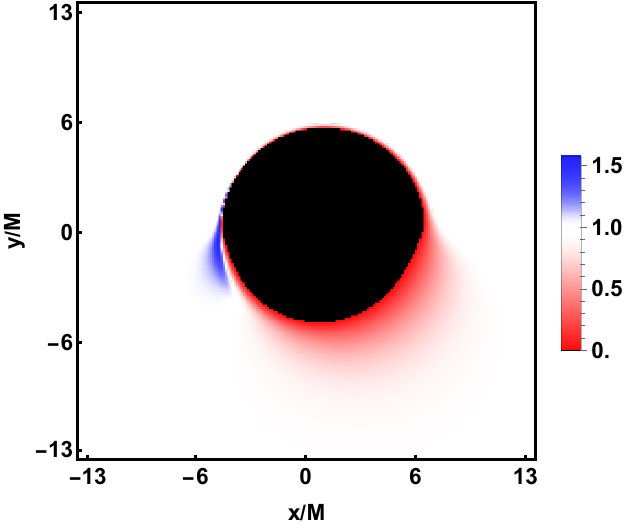}}
\subfigure[\tiny][~$Q=0.9,~\beta=2$]{\label{d13}\includegraphics[width=4cm,height=3.5cm]{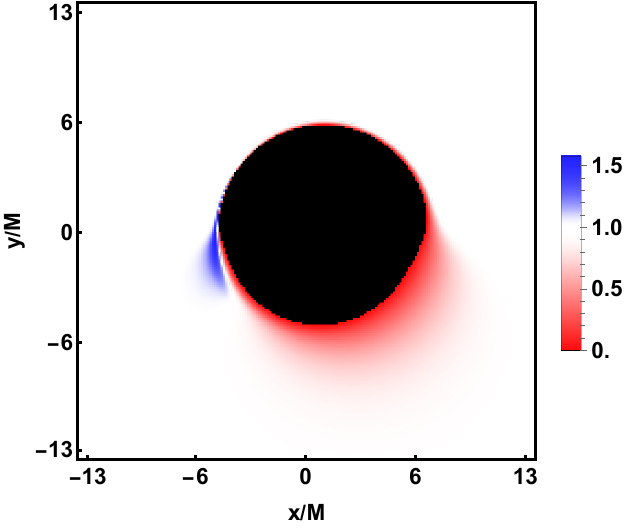}}
\caption{The redshift distribution of lensed images of the rotating BH in NED field is presented for different values of $Q$ and $\beta$, with the observer's inclination fixed at $\theta_{obs} = 80^\circ$, and $a=0.5$ under prograde accretion flow.}\label{prd7}
\end{figure}

Figure \textbf{\ref{prd7}} presents the redshift distribution of the lensed images for the rotating NED BH under the same parameter values used in the direct image analysis. In all panels, the lensed images are dominated by a redshifted emission surrounding the lower part of the BH shadow, whereas the blueshifted component appears only as a narrow and faint region on the opposite side of the image. With increasing $\beta$ from left to right, the redshifted emission extends over a slightly larger region of the image and follows the outward expansion of the shadow boundary, while the blueshifted region remains weak with only minor variations. On the other hand, increasing the $Q$ from the top to the bottom causes the central dark region to become more compact, accompanied by a corresponding contraction of the redshifted emission toward the shadow boundary. Throughout the considered parameter range, the lensed images are primarily characterized by redshifted radiation, whereas the contribution from the blueshifted emission remains comparatively insignificant.

\begin{figure}
\subfigure[\tiny][~$Q=0.1,~\beta=0.5$]{\label{a14}\includegraphics[width=4cm,height=3.8cm]{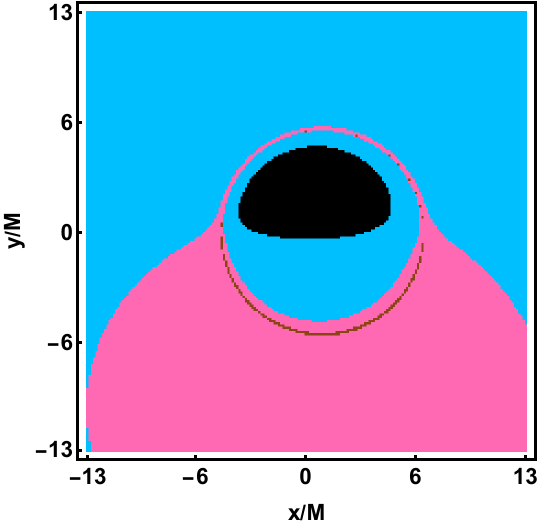}}
\subfigure[\tiny][~$Q=0.1,~\beta=1$]{\label{b14}\includegraphics[width=4cm,height=3.8cm]{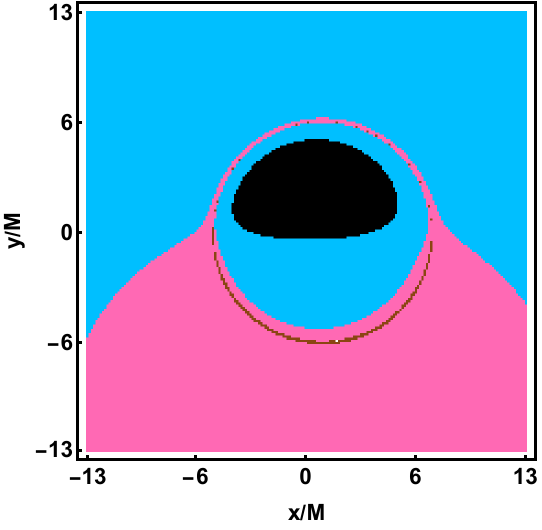}}
\subfigure[\tiny][~$Q=0.1,~\beta=1.5$]{\label{c14}\includegraphics[width=4cm,height=3.8cm]{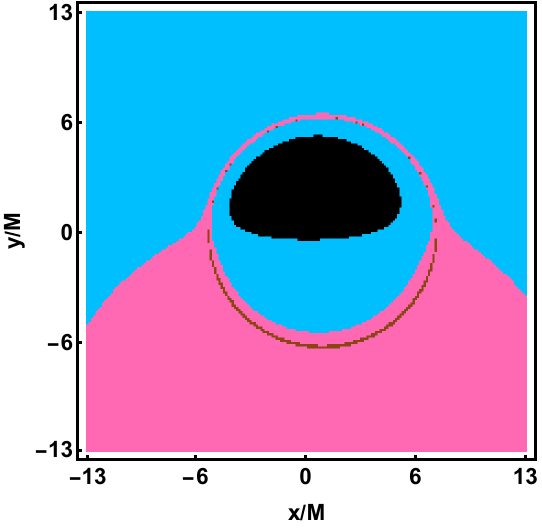}}
\subfigure[\tiny][~$Q=0.1,~\beta=2$]{\label{d14}\includegraphics[width=4cm,height=3.8cm]{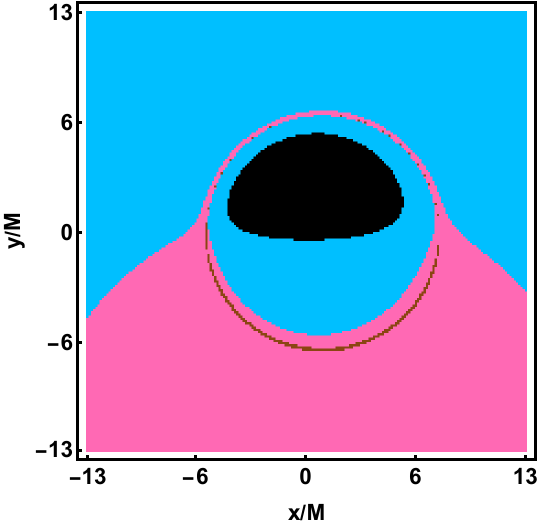}}
\subfigure[\tiny][~$Q=0.5,~\beta=0.5$]
{\label{a15}\includegraphics[width=4cm,height=3.8cm]{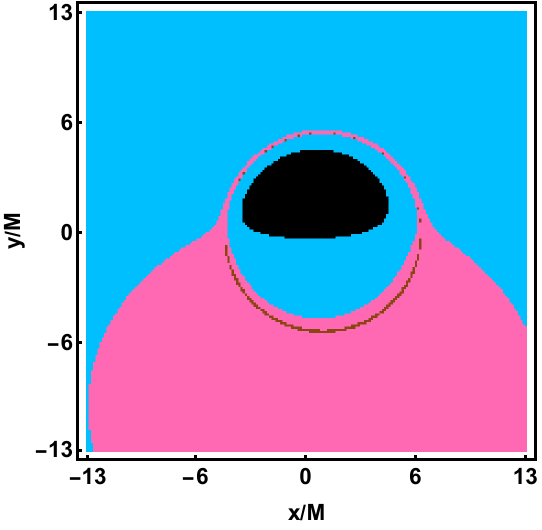}}
\subfigure[\tiny][~$Q=0.5,~\beta=1$]{\label{b15}\includegraphics[width=4cm,height=3.8cm]{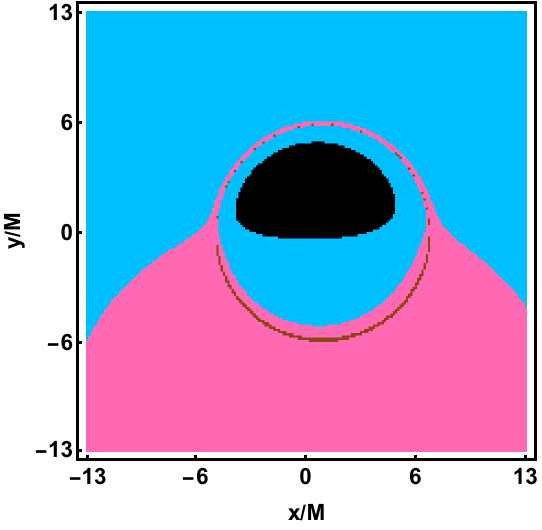}}
\subfigure[\tiny][~$Q=0.5,~\beta=1.5$]{\label{c15}\includegraphics[width=4cm,height=3.8cm]{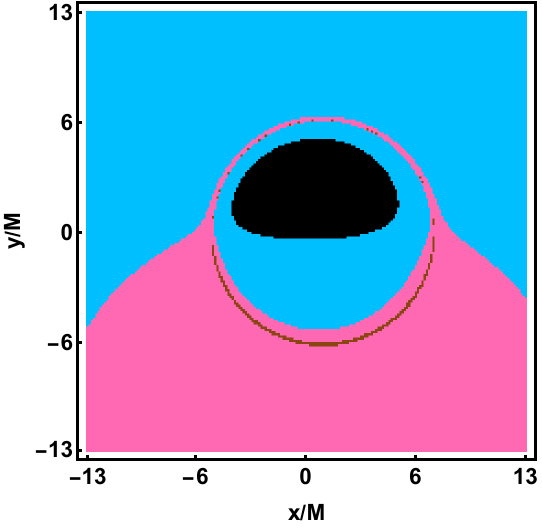}}
\subfigure[\tiny][~$Q=0.5,~\beta=2$]{\label{d15}\includegraphics[width=4cm,height=3.8cm]{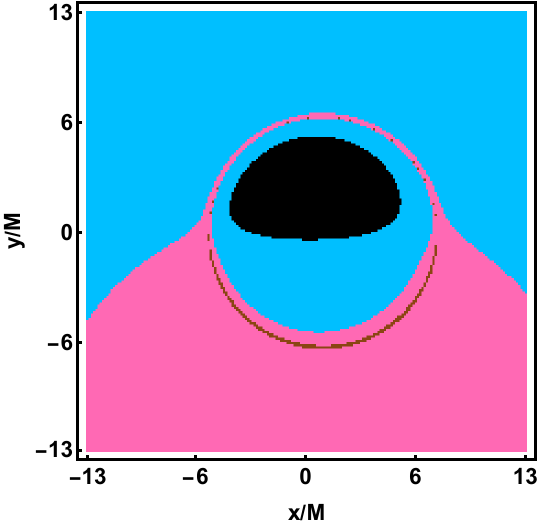}}
\subfigure[\tiny][~$Q=0.9,~\beta=0.5$]
{\label{a16}\includegraphics[width=4cm,height=3.8cm]{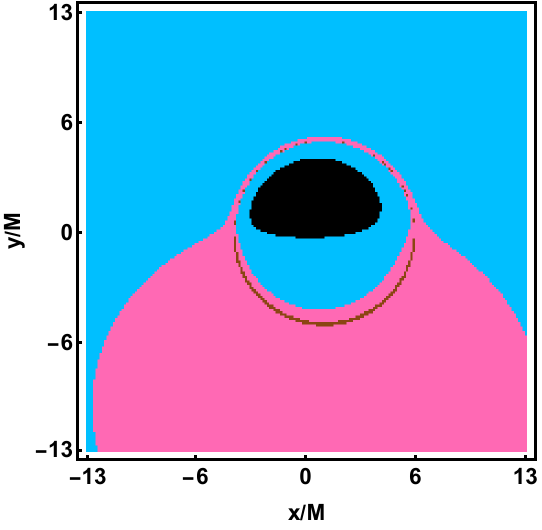}}
\subfigure[\tiny][~$Q=0.9,~\beta=1$]{\label{b16}\includegraphics[width=4cm,height=3.8cm]{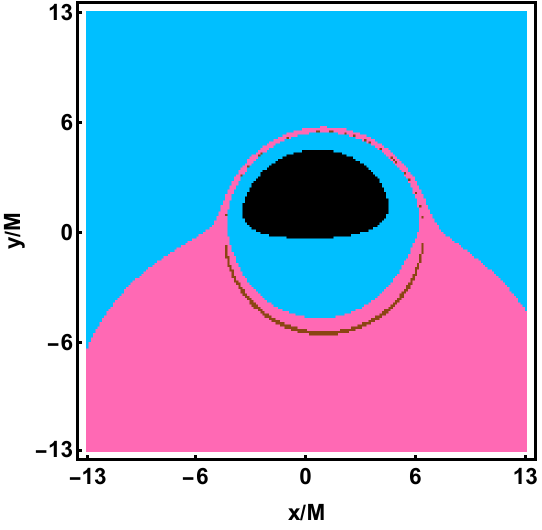}}
\subfigure[\tiny][~$Q=0.9,~\beta=1.5$]{\label{c16}\includegraphics[width=4cm,height=3.8cm]{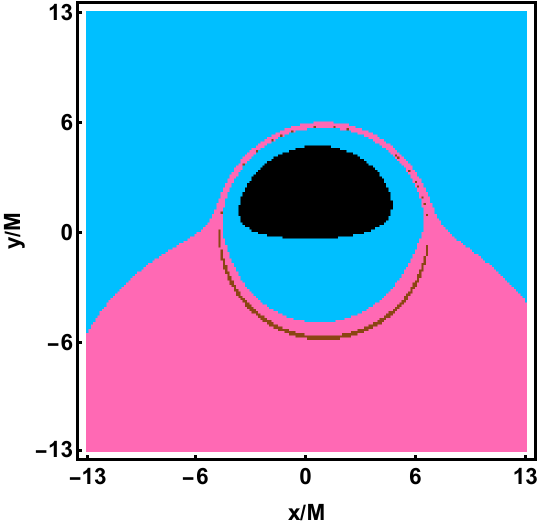}}
\subfigure[\tiny][~$Q=0.9,~\beta=2$]{\label{d16}\includegraphics[width=4cm,height=3.8cm]{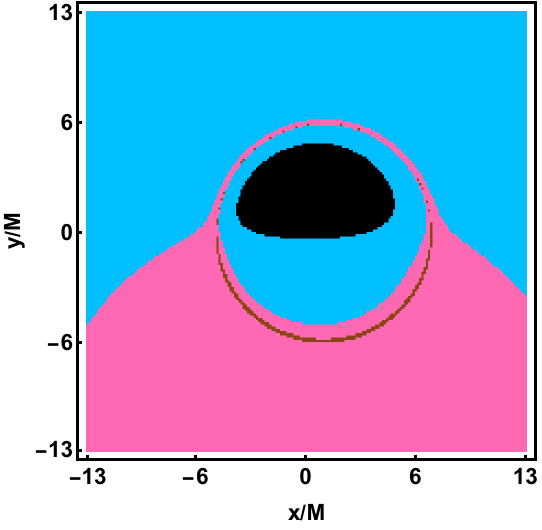}}
\caption{The lensing bands of the rotating BH in NED field is presented for different values of $Q$ and $\beta$, while keeping the observer's inclination fixed at $\theta_{obs} = 80^\circ$, and $a=0.5$ under prograde accretion flow.}\label{prd8}
\end{figure}

In Fig. \textbf{\ref{prd8}}, we present the direct and lensed emission bands of the rotating BH in NED field for different values of$Q$ and $\beta$. The sky blue and pink regions correspond to the direct and lensed emission bands, respectively, while the central black region represents the BH shadow enclosed by the photon ring. The photon ring clearly separates the two emission regions and remains visible in all panels. From left to right, the NED parameter increases from $0.5$ to $2$, whereas the electric charge is kept fixed in each row. As $\beta$ increases, the photon ring gradually shifts outward, accompanied by a slight expansion of both the direct and lensed emission bands. Another noticeable feature is the gradual change in the morphology of the central shadow. The upper part of the shadow becomes progressively flatter, giving rise to a more pronounced hat-like appearance for larger values of $\beta$. Consequently, the sky blue and pink regions occupy a relatively larger area around the photon ring, while the overall emission pattern remains nearly unchanged. The influence of the electric charge is displayed from the top to the bottom. As $Q$ increases, the central dark region gradually moves towards the upper side of the screen, and the photon ring moves closer to the BH center. Consequently, the direct and lensed emission bands shift inward, while the shadow boundary exhibits a slightly larger distortion. 
\begin{figure}
\subfigure[\tiny][~$Q=0.5,~\beta=0.5$]{\label{a17}\includegraphics[width=4cm,height=3.5cm]{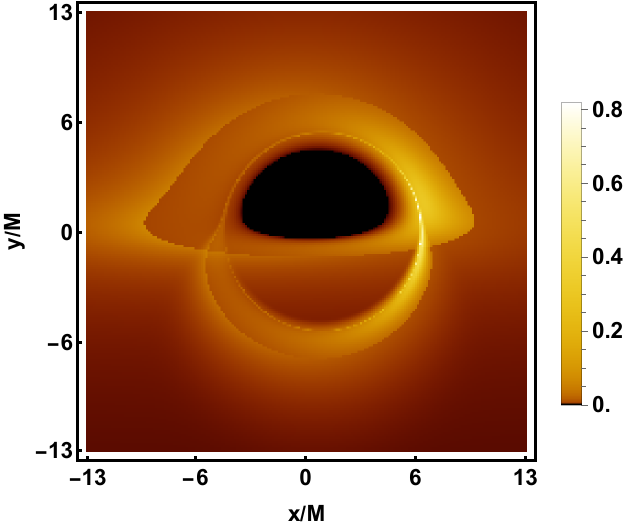}}
\subfigure[\tiny][~$Q=0.5,~\beta=1$]{\label{b17}\includegraphics[width=4cm,height=3.5cm]{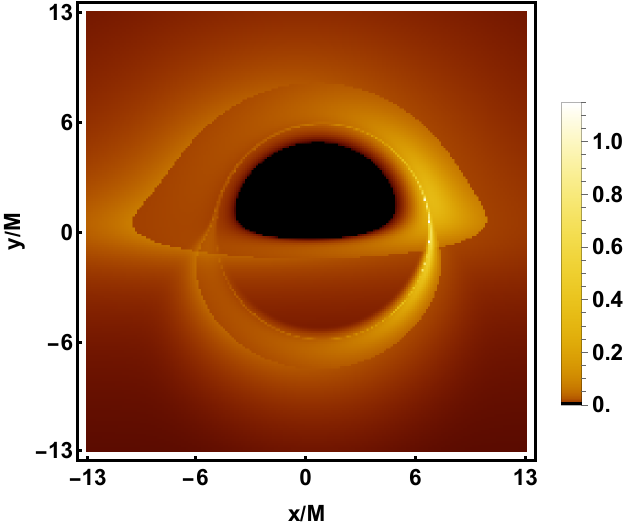}}
\subfigure[\tiny][~$Q=0.5,~\beta=1.5$]{\label{c17}\includegraphics[width=4cm,height=3.5cm]{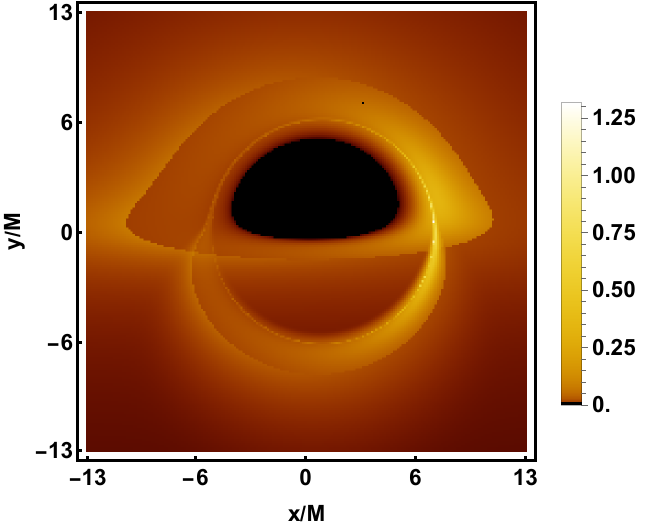}}
\subfigure[\tiny][~$Q=0.5,~\beta=2$]{\label{d17}\includegraphics[width=4cm,height=3.5cm]{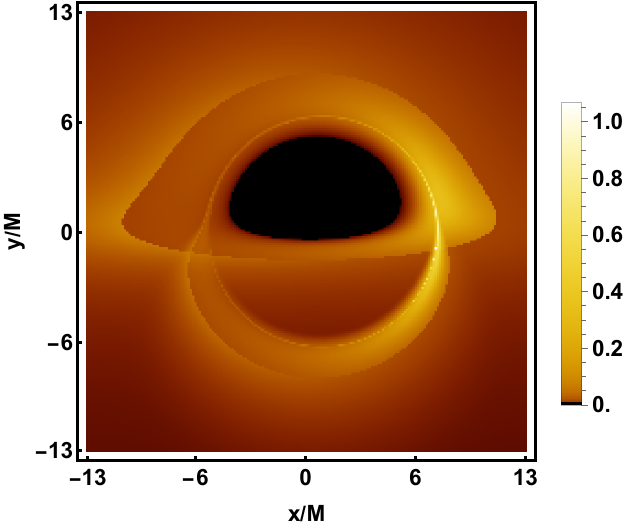}}
\caption{Optical images of the rotating BH in NED field under retrograde accretion flow for different values of $\beta$. The observer inclination is fixed at $\theta_{obs}=80^\circ$, $Q=0.5$ and $a=0.5$.}\label{prd9}
\end{figure}
Figure \textbf{\ref{prd9}} shows that the optical appearance of the rotating BH in NED field for the retrograde accretion flow with fixed $Q=0.5$, while the NED parameter $\beta$ increases from $0.5$ to $2$ from left to right. In all panels, the central black region represents the BH shadow, which has a clear hat-like shape. The shadow is surrounded by a bright photon ring, while the upper and lower bright regions correspond to the direct and lensed emission, respectively. The lower bright region is wider and brighter than the upper one, whereas the left side of the image appears narrower than the right side. As $\beta$ increases, the photon ring becomes slightly larger, and the hat-like shape of the shadow becomes more visible. The bright emission around the shadow also spreads slightly outward, while the overall appearance of the image remains almost the same.

\begin{figure}
\subfigure[\tiny][~$Q=0.5,~\beta=0.5$]{\label{a18}\includegraphics[width=4cm,height=3.5cm]{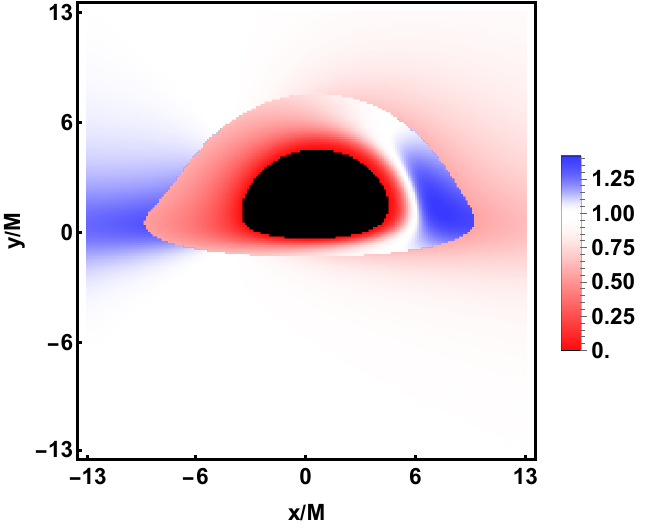}}
\subfigure[\tiny][~$Q=0.5,~\beta=1$]{\label{b18}\includegraphics[width=4cm,height=3.5cm]{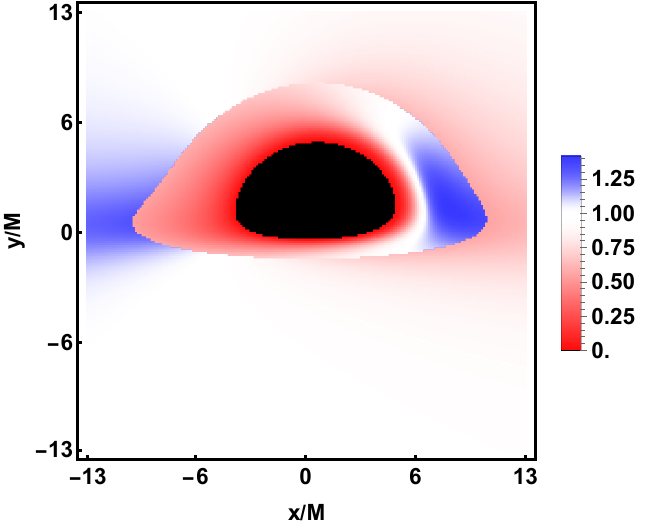}}
\subfigure[\tiny][~$Q=0.5,~\beta=1.5$]{\label{c18}\includegraphics[width=4cm,height=3.5cm]{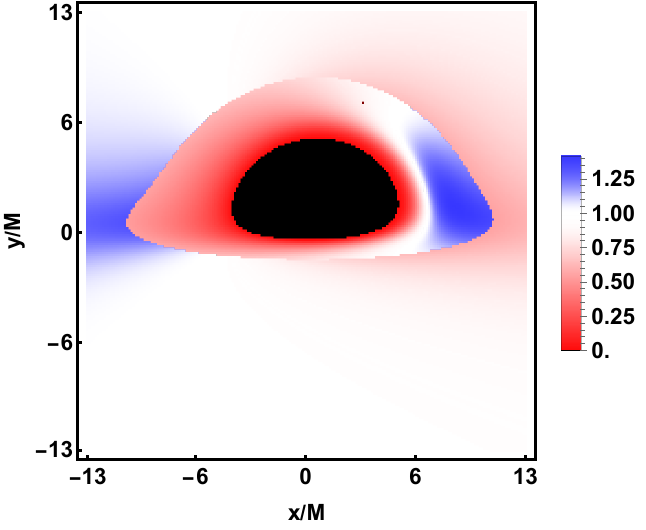}}
\subfigure[\tiny][~$Q=0.5,~\beta=2$]{\label{d18}\includegraphics[width=4cm,height=3.5cm]{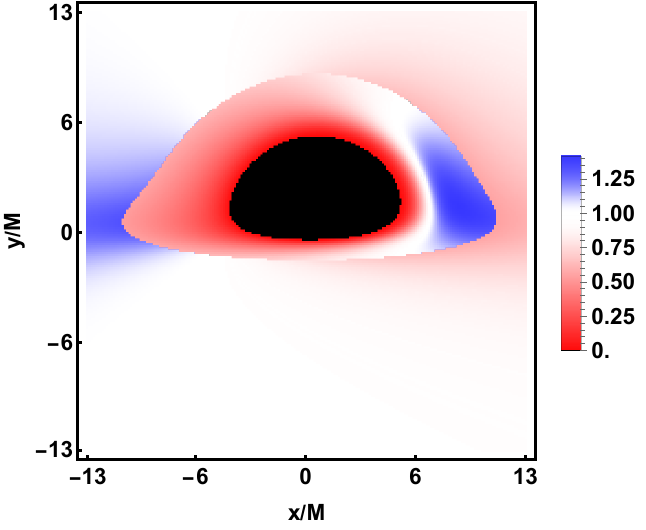}}
\caption{The redshift distribution of direct images for the rotating BH in NED field is presented for different values of $\beta$. The observer's inclination is fixed at
$\theta_{obs} = 80^\circ$, $Q=0.5$ and $a=0.5$ under retrograde accretion flow.}\label{prd10}
\end{figure}
Figure \textbf{\ref{prd10}} shows that the redshift distribution of the direct image for the retrograde accretion flow with $Q=0.5$, while the NED parameter increases from $0.5$ to $2$ from left to right. The central black region represents the BH shadow, which has a distinct hat-like shape in all panels. The upper part of the image is mainly covered by the redshifted emission, whereas the blueshifted region appears as a narrow crescent-shaped band on the right side of the shadow. The left side is almost entirely dominated by the redshifted emission. As $\beta$ increases, the shadow and the surrounding emission expand slightly, making the hat-like shape of the shadow more noticeable. The redshifted and blueshifted regions also become slightly broader, while their overall positions and shapes remain nearly unchanged. This indicates that increasing $\beta$ mainly affects the size of the observed image without significantly altering its redshift pattern.

\begin{figure}
\subfigure[\tiny][~$Q=0.5,~\beta=0.5$]{\label{a19}\includegraphics[width=4cm,height=3.5cm]{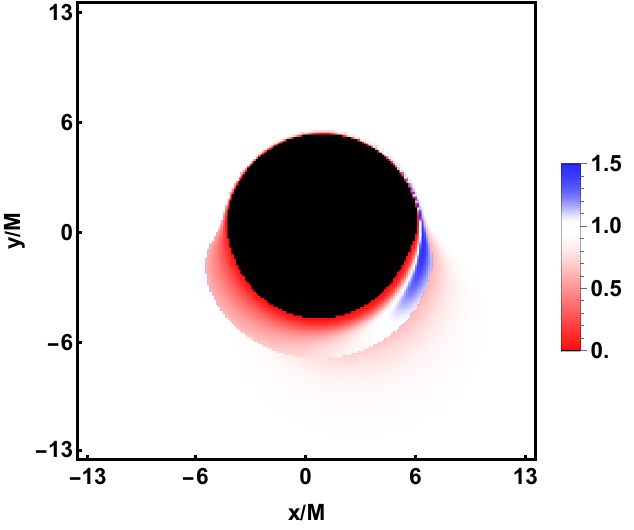}}
\subfigure[\tiny][~$Q=0.5,~\beta=1$]{\label{b19}\includegraphics[width=4cm,height=3.5cm]{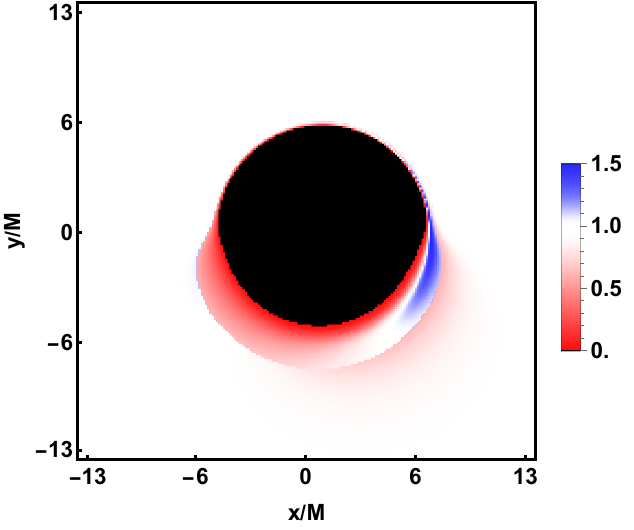}}
\subfigure[\tiny][~$Q=0.5,~\beta=1.5$]{\label{c19}\includegraphics[width=4cm,height=3.5cm]{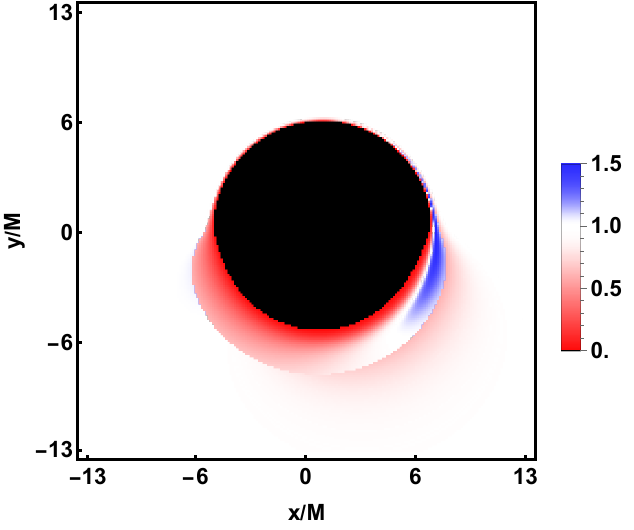}}
\subfigure[\tiny][~$Q=0.5,~\beta=2$]{\label{d19}\includegraphics[width=4cm,height=3.5cm]{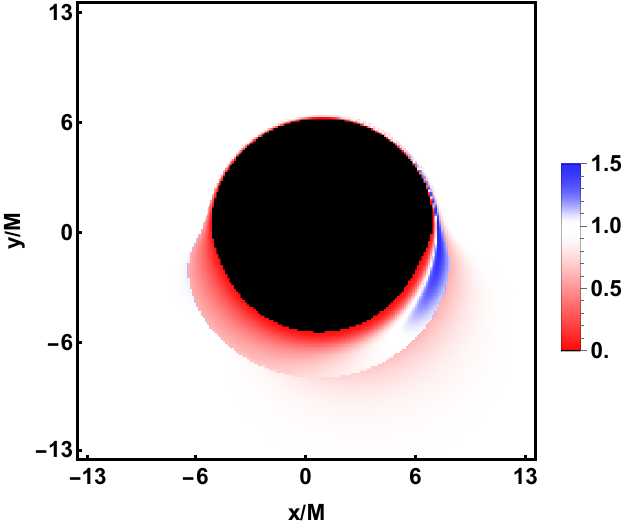}}
\caption{The redshift distribution of lensed images of the rotating BH in NED field is presented for different values of $\beta$. The observer's inclination is fixed at $\theta_{obs} = 80^\circ$, $Q=0.5$, and $a=0.5$ under retrograde accretion flow.}\label{prd11}
\end{figure}
The optical images of Fig. \textbf{\ref{prd11}} shows that the redshift distribution of the lensed image for the retrograde accretion flow with $Q=0.5$, while the NED parameter increases from $0.5$ to $2$ from left to right. The central black region represents the BH shadow, which remains nearly unchanged throughout all panels. The most noticeable changes occur below the shadow, where the redshifted emission forms a broad, bright region. As $\beta$ increases, this lower redshifted region gradually spreads over a larger area and extends farther away from the shadow boundary. On the opposite side, the blueshifted emission appears as a thin bright arc along the lower-right edge of the shadow and becomes slightly broader for larger values of $\beta$.

\begin{figure}
\subfigure[\tiny][~$Q=0.5,~\beta=0.5$]{\label{a20}\includegraphics[width=4cm,height=3.8cm]{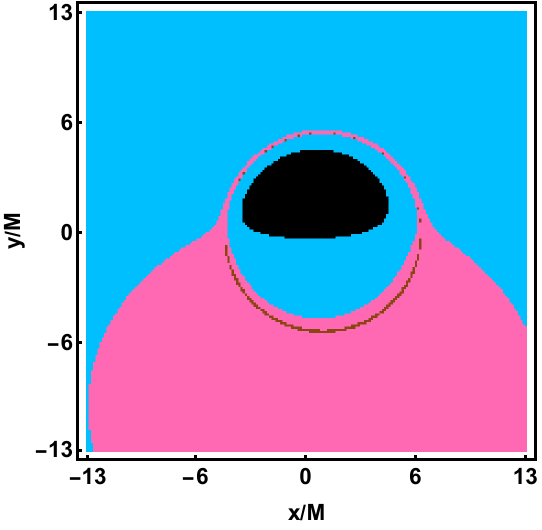}}
\subfigure[\tiny][~$Q=0.5,~\beta=1$]{\label{b20}\includegraphics[width=4cm,height=3.8cm]{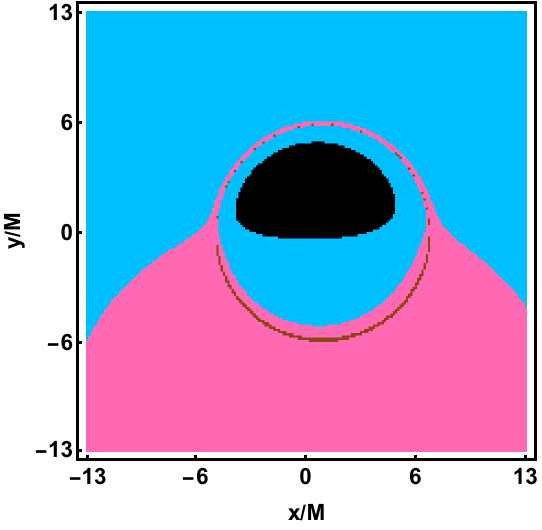}}
\subfigure[\tiny][~$Q=0.5,~\beta=1.5$]{\label{c20}\includegraphics[width=4cm,height=3.8cm]{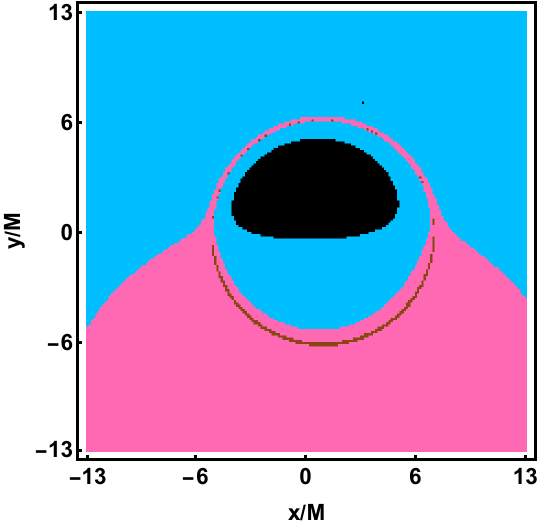}}
\subfigure[\tiny][~$Q=0.5,~\beta=2$]{\label{d20}\includegraphics[width=4cm,height=3.8cm]{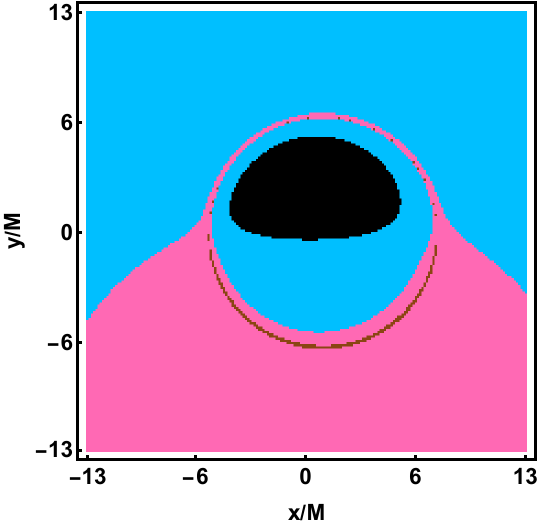}}
\caption{The lensing bands of the rotating BH in NED field are examined for different values of $\beta$. The observer's inclination is fixed at $\theta_{obs}=80^\circ$, $Q=0.5$, and $a=0.5$ under retrograde accretion flow.}\label{prd12}
\end{figure}
Figure \textbf{\ref{prd12}} shows the direct and lensed emission bands for the retrograde accretion flow with $Q=0.5$, while $\beta$ increases from $0.5$ to $2$ from left to right. The sky blue and pink regions denote the direct and lensed emission bands, respectively, whereas the central black region represents the BH shadow enclosed by the photon ring. As $\beta$ increases, the photon ring and the emission bands expand slightly, while the central shadow gradually develops a more pronounced hat-like shape. Consequently, the direct and lensed emission bands occupy a relatively larger area around the shadow. Despite these changes, the overall structure of the optical image remains nearly unchanged, with the photon ring clearly separating the direct and lensed emission bands in all panels.
\section{Results and Discussion}
In this work, we conducted an in-depth analysis of the optical characteristics of a rotating BH in the NED field, such as the geometrical shape of the shadow radius, visual signatures of shadow images in the background of a celestial light sphere and a thin accretion disk, which is consistent with Event Horizon Telescope results at $230$GHz. We examine the shadow observed by a distant observer and analyzed how the size of the shadow is influenced by the relevant parameters of the model. We impose the backward ray-tracing procedure together with a fisheye camera model, and observe the BH shadow images on the observer's screen. Particularly, we investigate the shadow observables, including the shadow radius and distortion parameter, as well as the direct and lensed images, redshift distributions, and emission bands for both prograde and retrograde accretion flows. The main results obtained from the analysis are summarized below. 

The horizon structure confirms the existence of the Cauchy and event horizons, whose locations vary with $\beta$. The shadow contours reveal that increasing $\beta$ enlarges the shadow and makes its boundary more circular, as shown in Fig. \textbf{\ref{prd2}}. The shadow observables further confirm that the NED and the electric charge have opposite effects on the shadow geometry. Increasing $\beta$ increases the shadow radius $R_d$ and decreases the distortion parameter $\delta_d$, whereas increasing $Q$ reduces $R_d$ and increases $\delta_d$ (see Fig. \textbf{\ref{prd3}}). The celestial sphere images in Fig. \textbf{\ref{prd4}} also exhibit noticeable variations with the model parameters. Larger values of $\beta$ produce a wider photon ring and broader lensing bands, while increasing $Q$ results in a more compact shadow and a more distorted optical appearance. For the prograde accretion flow, the optical images show that the NED parameter and the electric charge produce distinct effects on the appearance of the rotating NED BH. Larger values of $\beta$ lead to a wider photon ring and more extended direct and lensed emission bands, while the shadow becomes larger with a smoother boundary.
In contrast, increasing $Q$ makes the shadow more compact and increases its deformation, causing the emission features to move closer to the BH. The redshift analysis reveals that the observed images are mainly dominated by redshifted emission, whereas the blueshifted region remains confined to a small area near the photon ring. These results demonstrate that $\beta$ primarily controls the size of the optical image, while $Q$ has a stronger influence on its shape and asymmetry. For the retrograde accretion flow, the optical appearance exhibits a more asymmetric structure than in the prograde case. As $\beta$ increases, the shadow gradually develops a more pronounced hat-like shape, accompanied by a slight expansion of the photon ring and the surrounding emission regions. The direct and lensed images preserve their overall structure, although the lower-emission region becomes more extended as $\beta$ increases. The corresponding redshift maps show that the redshifted emission remains the dominant feature of the observed image. In contrast, the blueshifted emission is limited to a narrow region near the photon ring. These results indicate that the retrograde flow enhances the asymmetry of the optical image and makes the influence of the NED parameter more evident in the observed emission pattern.

 Overall, the present study provides a detailed analysis of the optical signatures of rotating BHs in NED field by combining shadow properties, celestial light sphere images, and thin accretion disk observations. The results demonstrate that the NED parameter and the electric charge leave distinct imprints on the horizon structure, shadow morphology, photon ring, and emission patterns, making these quantities useful for distinguishing rotating NED BHs from other BH spacetimes. With the rapid progress in BH imaging, these results may provide a useful framework for future comparisons with high-resolution observations and offer an additional avenue for testing NED in the strong gravity regime. As a possible extension of this work, it would be interesting to investigate polarized emission, time-dependent accretion flows, and magnetized plasma environments around rotating NED BHs. Such investigations could provide complementary observational constraints and improve our understanding of NED in astrophysical BH systems.

\end{document}